\input{psfig}
\documentclass[]{aa}

\usepackage{deluxetable}
\usepackage{aalongtable}
\begin{document}
\title{Long-term magnetic activity in close binary systems.\thanks{I dedicate this paper to the memory of the P.I. of this project, Prof. Marcello Rodon\`o, who suddenly passed away on October 23, 2005. To him my sincere estimation and deepest gratitude.}\\
I. Patterns of color variations\thanks{Based on
 observations collected at INAF-Catania Astrophysical Observatory, Italy.}\thanks{Tables 4-7 are available
 only in electronic form.}}
\author{Sergio Messina}
\offprints{Sergio Messina}
\institute{INAF-Catania Astrophysical Observatory, via S.\,Sofia 78, I-95123 Catania, Italy  \\
\email{sergio.messina@oact.inaf.it}\\
}

\date{}
\titlerunning{Patterns of color variations in close binary systems}
\authorrunning{S.\,Messina}

\abstract {}
 {This is the first of a series of papers in which we present the results of a long-term photometric monitoring project carried out at Catania Astrophysical Observatory and aimed at studying magnetic activity in late-type components of close binary systems, its dependence on global stellar parameters, and its evolution on different time scales from days to years. In this first paper, we present the complete observations dataset and new results of an investigation on the origin of brightness and color variations observed in the following well-known magnetically active close binary stars: \object{AR Psc}, \object{VY Ari}, \object{UX Ari}, \object{V711 Tau}, \object{EI Eri}, V1149 Ori, \object{DH Leo}, \object{HU Vir}, \object{RS CVn}, \object{V775 Her}, \object{AR Lac}, \object{SZ Psc}, \object{II Peg} and \object{BY Dra}} {About 38,000 high-precision photoelectric nightly observations in the U, B and V filters are analysed. Correlation and regression analyses of the V magnitude vs. U$-$B and B$-$V color variations are carried out and a comparison with model variations for a grid of active regions temperature and filling factor values is also performed.} {We find the existence of two different patterns of color variations. Eight stars in our sample: \object{BY Dra}, \object{VY Ari}, \object{V775 Her}, \object{II Peg}, V1149 Ori, \object{HU Vir}, \object{EI Eri} and \object{DH Leo} become redder when they get fainter, as it is expected from the presence of active regions consisting of cool spots. The other six stars show the opposite behaviour, i.e. they become bluer when they get fainter. For \object{V711 Tau} this behaviour could be explained by the increased relative U- and B- flux contribution by the earlier-type component of the binary system when the cooler component gets fainter. On the other hand, for  \object{AR Psc}, \object{UX Ari}, \object{RS CVn}, \object{SZ Psc} and \object{AR Lac} the existence of hot photospheric faculae must be necessarily invoked. We also found that in single-lined and double-lined binary stars in which the fainter component is inactive or much less active the V magnitude is correlated to B$-$V and U$-$B color variations in more than 60\% of observation seasons. The correlation is found in less than 40\% of observation seasons when the fainter component has a non-negligible level of activity and/or hot faculae are present but they are either spatially or temporally uncorrelated to spots.} {}


\maketitle
\rm
\section{Introduction}
\begin{table*}
\caption{Program stars: spectral type; rotation period; brightest V magnitude (V$_{\rm min}$), maximum light curve amplitude ($\Delta$V$_{\rm max}$), mean colors, and total flux ratios (L$_c$/L$_h$) in the V, B and U bands of the cool ($c$) to the hot ($h$) components}. \label{target} 
\begin{tabular}{l l l l@{\hspace{.1cm}} c  c c@{\hspace{.1cm}} c@{\hspace{.1cm}} c@{\hspace{.1cm}} c@{\hspace{.1cm}} c@{\hspace{.1cm}} c@{\hspace{.1cm}}}
\hline
 Program & HD  & Name  & Sp. Type  & Period  & V$_{\rm min}$ & $\Delta$V$_{\rm max}$ & $<B-V>$ & $<U-B>$ & \multicolumn{3}{c}{L$_c$/L$_h$}\\
Star  &  Number       &       &           &   (d)   &   (mag)       &    (mag)              & (mag)   & (mag)    & V  & B  & U \\
\hline
%
%
1  & 8357   & \object{AR Psc}     & K1 IV/V + G5/6 V    & 12.38      &  7.243    & 0.186  & 0.836 & 0.383 &5.6 & 4.8 & 3.5\\
2  & 17433  & \object{VY Ari}     & K3/4 IV + ?         & 16.3      &  6.690    & 0.421  & 0.979 & 0.649  &--- & --- & ---\\
3  & 21242  & \object{UX Ari}     & K0 IV + G5 V        &  6.43971   &  6.362    & 0.273  & 0.852 & 0.438 & 11.2 & 9.5 & 6.9\\ 
4  & 22468  & \object{V711 Tau}   & K1 IV + G5 V        &  2.83774   &  5.635    & 0.171  & 0.901 & 0.474 & 4.5  & 3.4  & 1.5\\
5  & 26337  & \object{EI Eri}     & G5 IV + G0 V        &  1.94722   &  6.921    & 0.159  & 0.662 & 0.105 &2.6 & 2.4 & 2.1\\
6  & 37824  & V1149 Ori  & K2/3 III + F8 V     & 53.12      &  6.593    & 0.277  & 1.155 & 0.963 & 14.8 & 8.1 & 2.5 \\ 
7  & 86590  & \object{DH Leo}     & K0 V + K7 V         & 1.070354   &  7.811    & 0.079  & 0.898 & 0.481 &6.5$^{\mathrm{a}}$ & 8.9$^{\mathrm{a}}$ & 20.8$^{\mathrm{a}}$\\
8  & 106225 & \object{HU Vir}     & K1 IV + ?           & 10.41   &  8.549    & 0.419  & 1.023 & 0.647 &--- & --- & ---\\ 
9  & 114519 & \object{RS CVn}     & K0 IV + F5 V      & 4.797855   &  7.858    & 0.203  & 0.621 & 0.103 &0.9 & 0.5 & 0.3\\
10  & 175742 & \object{V775 Her}   & K0 V + K5/M2 V     & 2.879342   &  7.800    & 0.185  & 0.904 & 0.609 & 15.8$^{\mathrm{a}}$ & 22.2$^{\mathrm{a}}$ & 53.0$^{\mathrm{a}}$\\
11 & 210334 & \object{AR Lac}     &  K2 IV  + G2 IV      & 1.98322195 &  6.030    & 0.160  & 0.768 & 0.725 &1.4 & 1.2 & 0.9\\
12 & 219113 & \object{SZ Psc}     &  K1 IV + F8 V/IV     & 3.9657889  &  7.155    & 0.115   & 0.846 & 0.365 &3.8 & 2.8 & 1.7\\ 
13 & 224085 & \object{II Peg}     & K2 IV + ?           & 6.720      &  7.283    & 0.671  & 1.031 & 0.761   &--- & --- & ---\\ 
14 & 234677 & \object{BY Dra}     & M0 V + M0 V         & 3.836      &  8.003    & 0.176  & 1.172 & 1.043 & 1.0 & 1.0 & 1.0\\
\hline
\end{tabular}
\begin{list}{}{}
\item[$^{\mathrm{a}}$] total flux ratio of the hot to the cool component
\end{list}
\end{table*}
\begin{figure}
\begin{minipage}{10cm}
\psfig{file=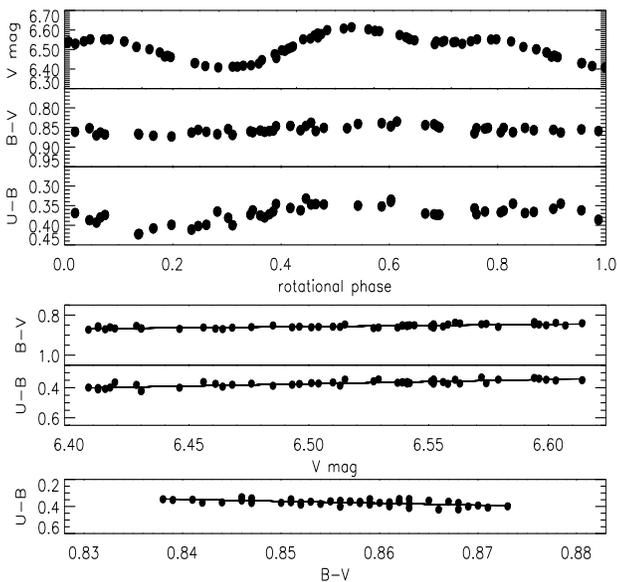,width=9cm,height=8cm}
\end{minipage}
\caption{\label{uxari_curva} \it Upper panels: \rm V magnitude, B$-$V and U$-$B colors  of \object{UX Ari} vs. rotational phase for the mean epoch 1993.82. \it Lower panels: \rm  Color-magnitude (B$-$V and U$-$B vs. V) and color-color (U$-$B vs. B$-$V) variations along with a linear fit (solid line). }
\end{figure}
Studies of stellar magnetic activity and variability at Catania Astrophysical Observatory (OAC)
date back to the late Sixties, when a pioneering research was undertaken to explore the nature of stellar spots 
(Rodon\`o \cite{Rodono65}) and stellar 
flares (Cristaldi \& Rodon\`o \cite{Cristaldi68}; Cristaldi, Narbone \& Rodon\`o \cite{Cristaldi_et_al68}).\\
\indent
The existence of stellar spots, first proposed by Kron (\cite{Kron50}) as cause of the variability observed in a few late-type stars, was still debated at that time. The research at OAC 
significantly contributed to understand their nature and yielded relevant results such as the discovery 
of the characteristic outside-of-eclipse \it photometric \rm or \it distortion wave \rm  on the light curve of the proto-type \object{RS CVn} system 
(Chisari \& Lacona \cite{Chisari65}; Catalano \& Rodon\`o \cite{Catalano67}) playing an important role in the identification of the new class 
of binaries named after \it \object{RS CVn} \rm (Oliver \cite{Oliver74}; Hall \cite{Hall76}).
Also stellar flare studies at OAC, contemporary carried out within the coordination of 
the \it Working Group on Flare Stars \rm 
(Gershberg \& Shakhovskaya \cite{Gershberg03}), yielded relevant progresses and had dedicated the 71st IAU
Colloquium on \it Activity in Red Dwarf Stars \rm  (Byrne \& Rodon\`o \cite{Byrne83}). \\
\indent
Research on stellar spots, during the course of many years, has progressively revealed that spots are non-stationary phenomena that after a few stellar
rotations undergo, depending on the value of global stellar properties, sizeable changes in their dimension, number and surface distribution. Furthermore, if the star rotates differentially, spots at different latitudes produce different variability terms in the frequency domain and any initial spot distribution is significantly sheared after some rotations (Lanza et al. \cite{Lanza93}; \cite{Lanza94}). 
Since light curves undergo noticeable changes, active stars must be monitored as continuously as possible if we want to derive significant information on the behaviour of stellar activity. For this reason, the photometric monitoring at OAC  became more and more systematic and was extended during the last three decades to a much larger sample of stars, either single or in close binary systems with a wide range of values of global properties (see, e.g., the series of papers by Cutispoto \cite{Cutispoto90}; \cite{Cutispoto91}; \cite{Cutispoto92}; \cite{Cutispoto93}; \cite{Cutispoto95}). The photometric patrol, initially carried out with the 30-cm and 91-cm telescopes of OAC, was afterwards continued mainly with the use of two APTs, i.e. automatic photometric telescopes: the \it Phoenix-25 \rm since 1988 (Rodon\`o \& Cutispoto \cite{Rodono92}) and the Catania \it APT80/1\rm, entirely dedicated to this project since 1992 (Rodon\`o et al. \cite{Rodono01b}; Messina, Rodon\`o \& Cutispoto \cite{Messina04}). Indeed, very high duty cycle and fully automation have revealed the APTs to be best suited to obtain a homogeneous and systematic data base of high-quality multiband photometry of magnetically active stars (Rodon\`o \cite{Rodono92}; Rodon\`o \& Cutispoto \cite{Rodono94a}; \cite{Rodono94b}).\\
\indent
Starting from the Eighties, other institutions begun similar long-term photometric monitoring projects with the use of robotic telescopes. Among the most relevant projects, we just mention the \it Sun in Time \rm undertaken by the Villanova University (see, e.g., Messina \& Guinan \cite{Messina02}; \cite{Messina_guinan03}). All these projects have made feasible a direct comparison between theoretical predictions and observational results, contributing significantly to increase our knowledge on starspot properties, active region growth and decay (ARGD), activity cycles, surface differential rotation, active longitudes and flip-flop phenomena, orbital period variations and their dependence on stellar parameters. \\
\indent
In a series of papers we will present the final results on active close binary systems obtained by the long-term photometric monitoring  project of OAC. For instance, the results of a similar project, but on single main-sequence stars,
was presented in a previous series of papers (Messina \& Guinan \cite{Messina02}; \cite{Messina_guinan03}).
In this first paper we investigate the origin of different patterns of color variations shown by active close binary systems. Starspots cycles and surface differential rotation will be the main subjects of following papers.\\
\indent
 The stellar sample and the photometric database are presented in Sect. 2. and 3.
In Sect. 4 we investigate the correlation between color and magnitude variations on both short and 
long timescales. In Sect. 5 we describe a simple modelling approach to probe the nature of color variations.  Discussion and conclusions are given in Sect. 6 and 7, respectively.

\begin{figure*}
\begin{minipage}{18cm}
\centerline{
\psfig{file=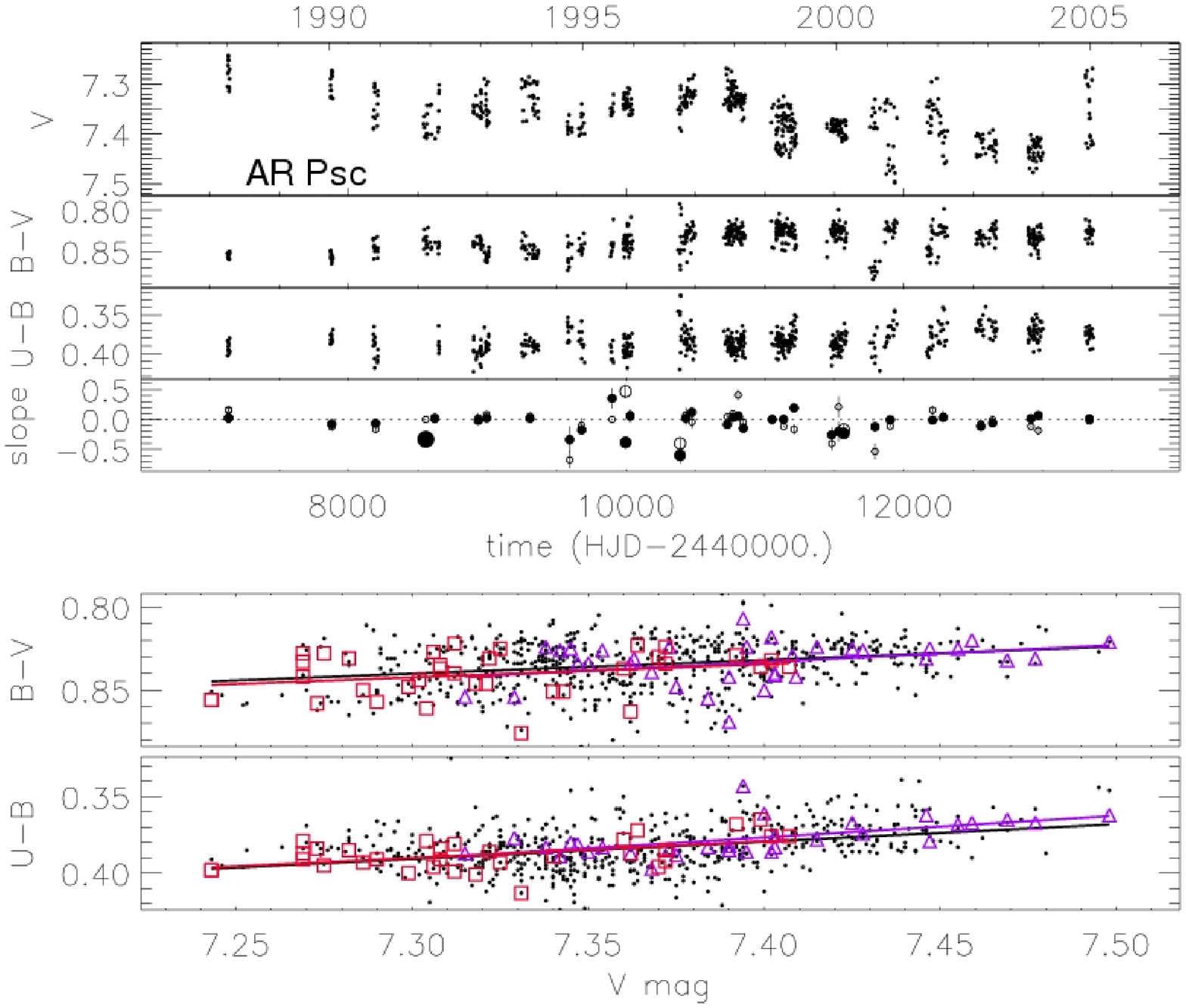,width=10cm,height=10cm,angle=90}
\psfig{file=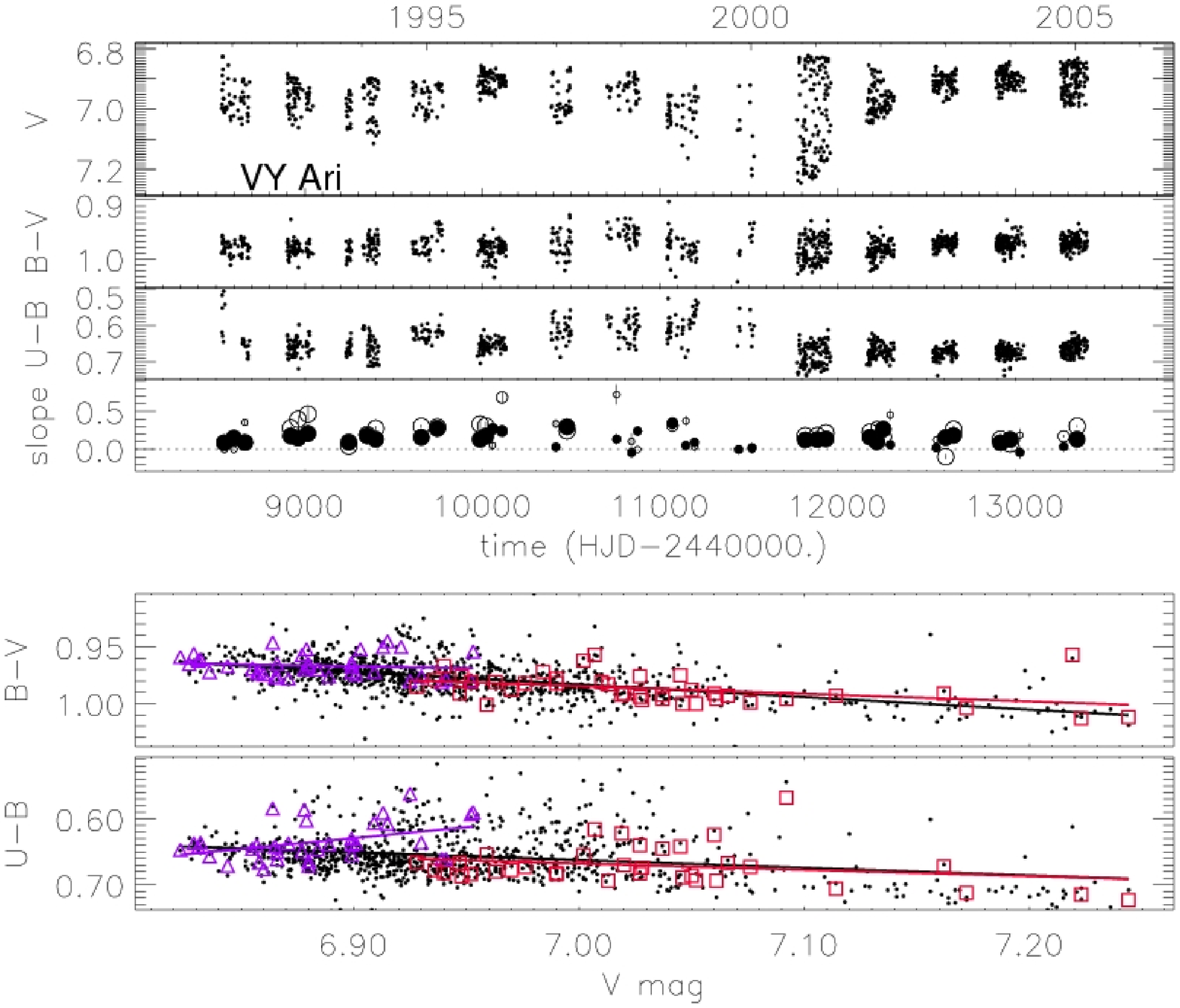,width=10cm,height=10cm,angle=90}
}
\end{minipage}
\caption{\label{fig-arpsc} {\it Left plot:} From top to bottom  V-band magnitudes, B$-$V and U$-$B colors, and slopes of the linear fit to the B$-$V vs. V (filled circles) and  U$-$B vs. V (open circles) relations vs. time for \object{AR Psc}. Larger symbol size indicates a larger significance level.  Bottom panels:  B$-$V and U$-$B colors vs. V magnitudes (dots). Triangles and squares show the relation between bluest and brightest and between reddest and faintest light curve values, respectively. Solid lines are linear fits to the mentioned relations. {\it Right plot:} the same as in the left plot, but for \object{VY Ari}.}
\end{figure*}
 \begin{figure*}
\begin{minipage}{18cm}
\centerline{
\psfig{file=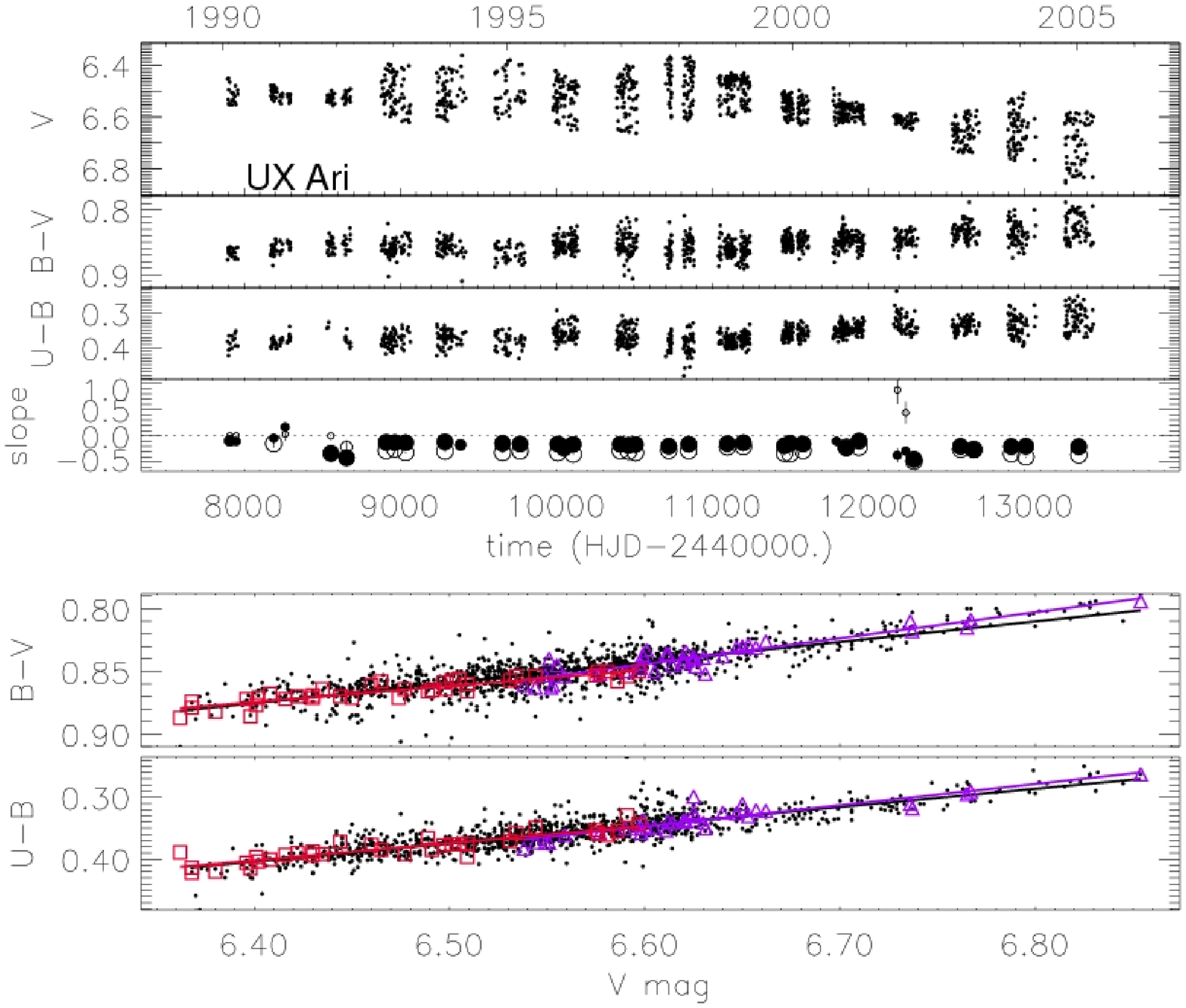,width=10cm,height=10cm,angle=90}
\psfig{file=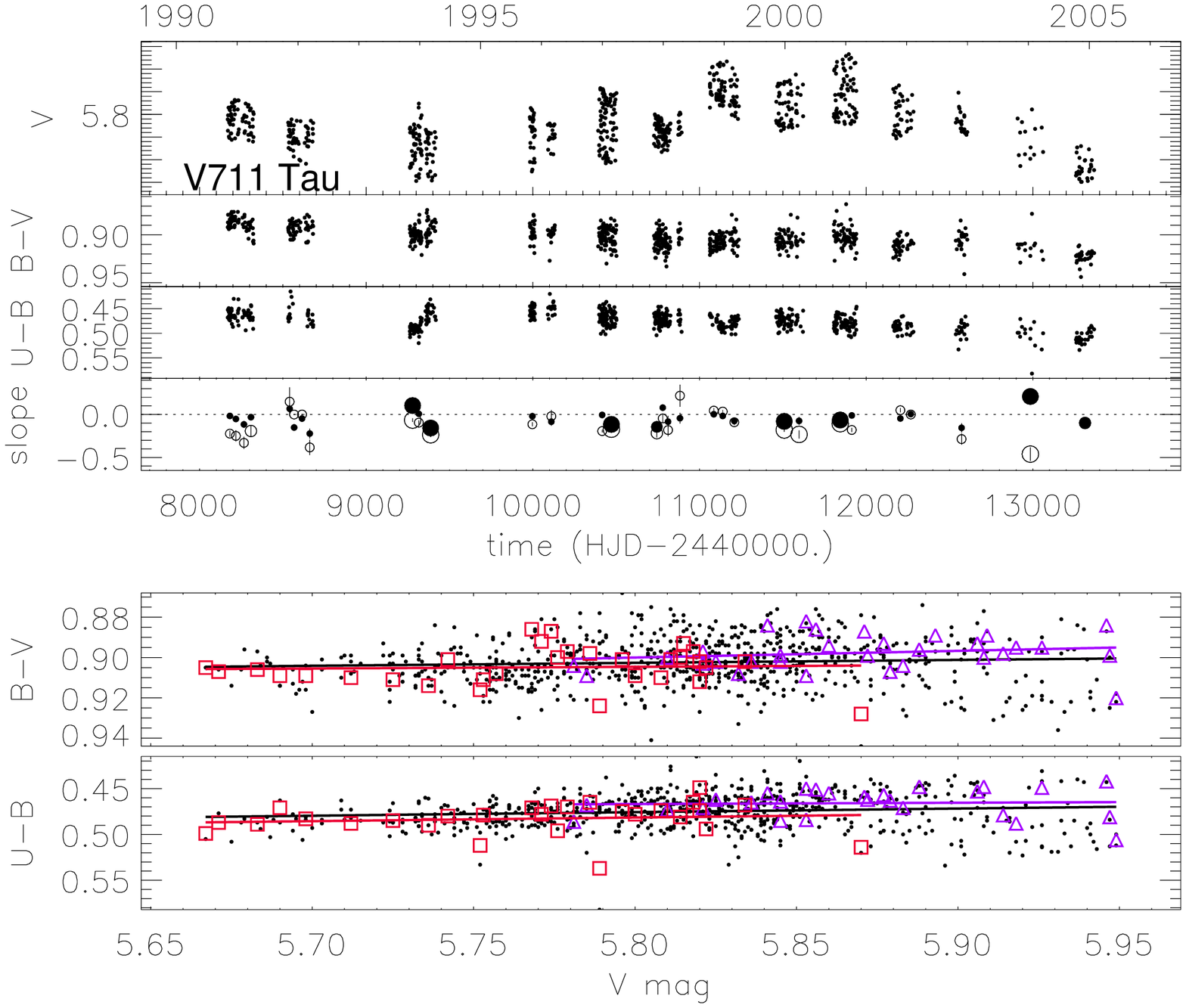,width=10cm,height=10cm,angle=90}
}
\end{minipage}
\caption{ As in Fig.~\ref{fig-arpsc}, but for \object{UX Ari} and \object{V711 Tau}. }
\end{figure*}
\begin{figure*}
\begin{minipage}{18cm}
\centerline{
\psfig{file=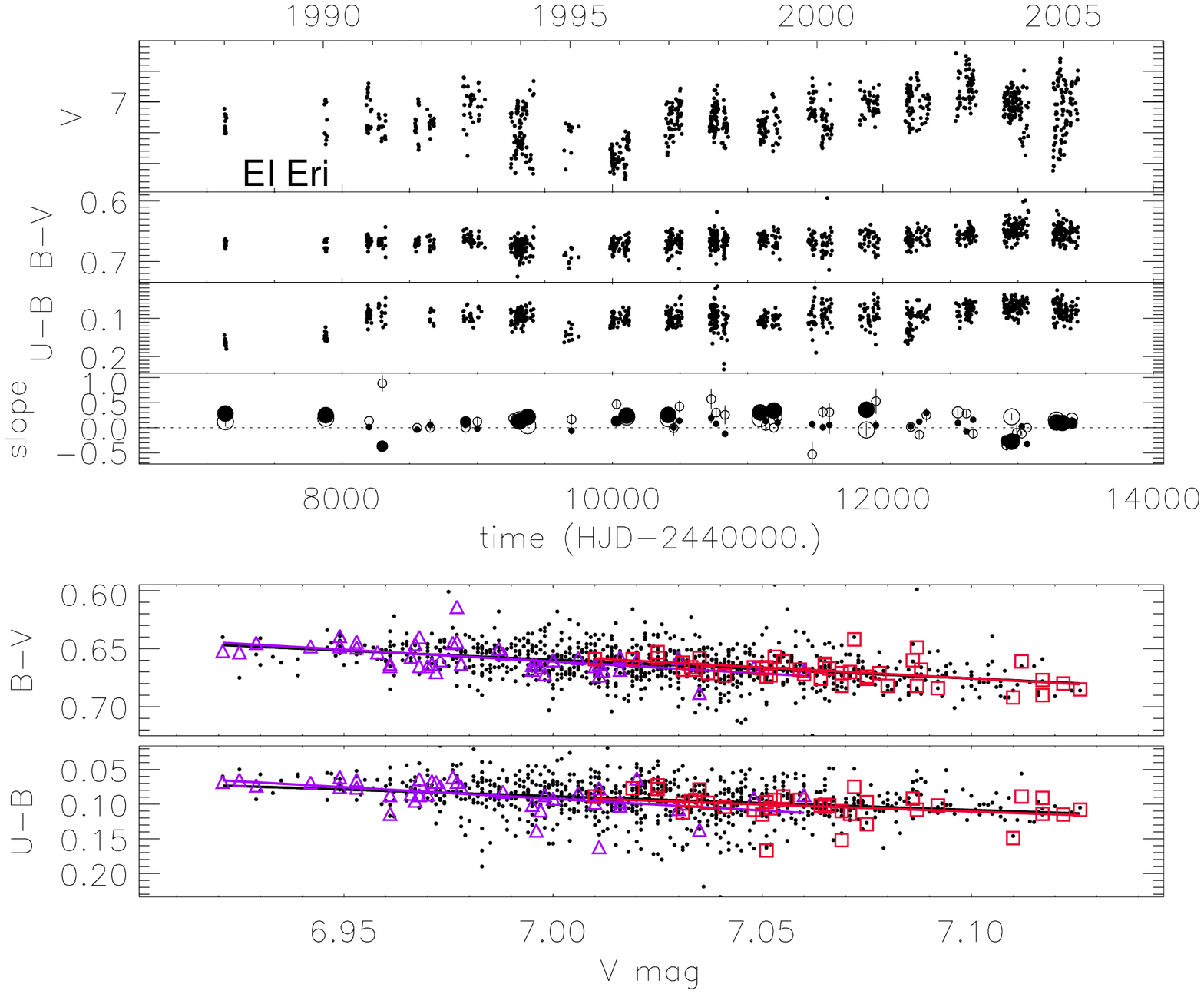,width=10cm,height=10cm,angle=90}
\psfig{file=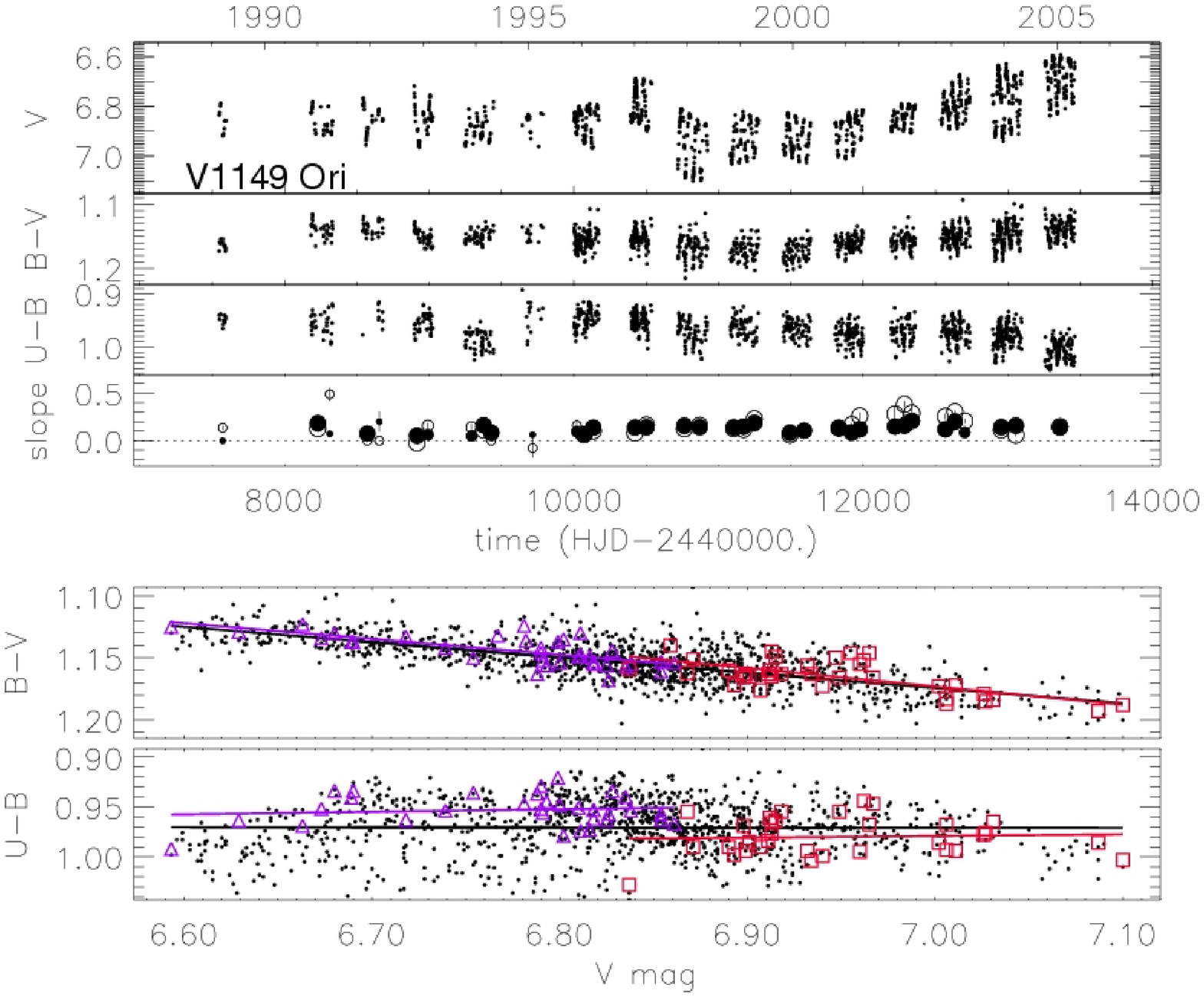,width=10cm,height=10cm,angle=90}
}
\end{minipage}
\caption{ As in Fig.~\ref{fig-arpsc}, but for \object{EI Eri} and V1149 Ori. }
\end{figure*}
\section{The stellar sample}
From the whole stellar sample of about fifty stars monitored at OAC (see Rodon\`o et al. \cite{Rodono01b}), we selected for the present analysis those stars for which we obtained the most extended data time series. The stellar sample analysed in this paper consists of 14 binary systems: six are SB1-type systems; eight are SB2-type systems, three of which are detached eclipsing binaries (see Table\,\ref{target}). 
In this Section we give some information from the literature on their optical behaviour and the values of their physical parameters that will be used in the following to model the color variations. In the following we indicate as primary of the binary system the more massive and luminous component, not of earlier spectral type.\\
\indent 
{\bf \object{AR Psc} (HD\,8357)} is an SB2 \object{RS CVn}-type variable consisting of a primary K1 subgiant and a secondary G5/6 dwarf (Cutispoto \cite{Cutispoto98}). This system has an orbital period of P$_{\rm orb} = 14.3023^d$ (Fekel \cite{Fekel96}), which is not synchronized with the rotational photometric period of P$_{\rm pho} = 12.38^d$ (Cutispoto, Messina \& Rodon\`o \cite{Cutispoto01}). The observed photometric variability is likely dominated by the more active and luminous subgiant component. However, the anticorrelation between U$-$B and V data found by Cutispoto (\cite{Cutispoto95}) may arise from the earlier-type component, as it will be discussed. We adopt for the hot ($h$) and cool ($c$) components the following values from Strassmeier et al. (\cite{Strassmeier93}), Hongguam et al. (\cite{Hongguam06}), and Cox (\cite{Cox00}): T$_h = 5600$\,K, T$_c = 4880$\,K, R$_h = 0.92$ R$_{\odot}$, R$_c = 2.7$ R$_{\odot}$, log\,$g_h$ = 4.5, log\,$g_c$ = 3.0 cm s$^{-2}$, M$_h$/M$_c$ = 0.82, $i$ = 37$^{\circ}$. By using Eqs.\,(1) and (6) from Morris \& Naftilan (\cite{Morris93}, and references therein), we computed that the variability of \object{AR Psc} arising from proximity and reflection effects is about $\Delta$V$_{\rm ellip}= 6.0\cdot 10^{-4}$ mag and   $\Delta$V$_{\rm ref}=1.3\cdot 10^{-4}$ mag.\\
\indent 
{\bf \object{VY Ari} (HD\,17433)} is an SB1 \object{RS CVn}-type variable with a primary K3/4 subgiant. This system has an orbital period of P$_{\rm orb} = 16.42^d$ (Strassmeier \& Bopp \cite{Strassmeier92}), and a variable rotational photometric period of about P$_{\rm pho} \sim 16.3^d$ (Messina, Rodon\`o \& Cutispoto \cite{Messina04}; Frasca et al. \cite{Frasca05}). On the basis of the Li 6707.8 \AA\,\, abundance, a significant infrared excess and the non-synchronized orbital/rotational period, Bopp et al. (\cite{Bopp89}) suggest that this star could be a PMS system. We adopt the effective temperature from Frasca et al. (\cite{Frasca05}) and the surface gravity  proper for its spectral class (Cox \cite{Cox00}): T$_h = 4900$\,K, log\,$g_h$ = 4.0 cm s$^{-2}$.  \object{VY Ari} will be treated in the analysis as a single star and proximity and reflection effects will be considered negligible.\\
\indent 
{\bf \object{UX Ari} (HD\,21242)} is a triple system consisting of a SB2 \object{RS CVn}-type binary having a primary K0 subgiant and a secondary G5 dwarf (Cutispoto, Messina \& Rodon\`o \cite{Cutispoto01}), and of a third faint late-type  star (Duemmler \& Aarum \cite{Duemmler01}). This system has a rotational photometric period similar to the orbital period of P$_{\rm orb} = 6.43791^d$ (Carlos \& Popper \cite{Carlos71}). \object{UX Ari} is known to have B$-$V and U$-$B color variations anticorrelated with the V-band flux variation, i.e. when the star becomes fainter it gets bluer (Zeilik et al. \cite{Zeilik82}; Rodon\`o \& Cutispoto \cite{Rodono92}). A flux contribution by the earlier-type component has been suspected as the cause of the color blueing (Wacker et al. \cite{Wacker86}; Mohin \& Raveendran \cite{Mohin89}). We adopt for the system's components the values of parameters from Aarum Ulvas \& Engvold (\cite{Aarum03}) and Strassmeier et al. (\cite{Strassmeier93}): T$_h = 5620$\,K, T$_c = 4750$\,K, T$_3 = 4400$\,K, R$_h = 1.11$ R$_{\odot}$, R$_c = 5.78$ R$_{\odot}$, R$_3 = 0.70$ R$_{\odot}$, M$_h$/M$_c$ = 0.80, $i$ = 60$^{\circ}$, and the surface gravity proper for their spectral classes: log\,$g_h$ = 4.5, log\,$g_c$ = 3.5, log\,$g_3$ = 4.5 cm s$^{-2}$. The flux from the third component has been properly scaled to take into account the parallax difference with respect to the K0\,IV + G5\,V binary. We find that the variability of \object{UX Ari} arising from proximity and reflection effects is about $\Delta$V$_{\rm ellip}= 8.7\cdot 10^{-3}$ mag and   $\Delta$V$_{\rm ref}=4.3\cdot 10^{-3}$  mag (Morris \& Naftilan \cite{Morris93}). \\
\indent 
{\bf \object{V711 Tau} (HD\,22468)} is an SB2 \object{RS CVn}-type variable consisting of a primary K1 subgiant and a secondary G5 dwarf (Fekel \cite{Fekel83}). The \object{RS CVn}-type binary belongs to a visual binary, ADS\,2644B (K3V) being its secondary component. The system has an orbital period of P$_{\rm orb} = 2.83774^d$ (Fekel \cite{Fekel83}) which does not differ significantly from the photometric rotational period (Lanza et al. \cite{Lanza06}). Although the observed photometric variability is dominated by the more active and luminous subgiant component, a flux contribution to the B and U bands by the earlier-type component has been suspected to be the cause of the color blueing (Aarum Ulvas \& Henry \cite{Aarum05}). We adopt the parameter values for the hot and cool components from Lanza et al. (\cite{Lanza06}) and for the third component from Aarum Ulvas \& Engvold (\cite{Aarum03}): T$_h = 5500$\,K, T$_c = 4750$\,K, T$_3 = 4950$\,K, R$_h = 1.1$ R$_{\odot}$, R$_c = 3.7$ R$_{\odot}$, R$_3 = 0.70$ R$_{\odot}$, M$_h$/M$_c$ = 0.80, $i$ = 38$^{\circ}$, and log\,$g_h$ = 4.26, log\,$g_c$ = 3.3, log\,$g_3$ = 4.5 cm s$^{-2}$. We find that the variability of \object{V711 Tau} arising from proximity and reflection effects is about $\Delta$V$_{\rm ellip}= 0.040$  mag and   $\Delta$V$_{\rm ref}=3.3\cdot 10^{-3}$ mag (Morris \& Naftilan \cite{Morris93}). For this stars Henry et al.  (\cite{Henry95}) established  that  the proximity effect gives a much smaller contribution of about $\Delta$V$_{\rm ellip}= 0.017$.\\
\indent 
{\bf \object{EI Eri} (HD\,26337)} is an SB1 \object{RS CVn}-type variable with a primary G5 subgiant and colors from UBVRI photometry consistent with a G5 IV + G0 V binary system (Cutispoto \cite{Cutispoto95}). \object{EI Eri} has an orbital period of   P$_{\rm orb} = 1.94722^d$ (Fekel et al. \cite{Fekel87}) and a variable rotational photometric period of about  P$_{\rm pho} = 1.94^d$ (Strassmeier et al. \cite{Strassmeier97}). If granted the spectral classification by Cutispoto (\cite{Cutispoto95}), the observed photometric variability may partly originate from the less-active main-sequence component, as will be discussed in the following. Evidence for an anticorrelation between U$-$B and V data were found by Rodon\`o \& Cutispoto (\cite{Rodono92}). We adopt for the hot and cool components the values of parameters proper for their spectral classes: T$_h = 5900$\,K, T$_c = 5600$\,K, R$_h = 1.1$ R$_{\odot}$, R$_c = 2.0$ R$_{\odot}$, M$_h$/M$_c$ $\simeq$ 1.0, $i$ = 50$^{\circ}$ (Strassmeier et al. \cite{Strassmeier93}), and log\,$g_h$ = 4.5, log\,$g_c$ = 3.5 cm s$^{-2}$ (Cox \cite{Cox00}). We find that the variability of \object{EI Eri} arising from proximity and reflection effects is about $\Delta$V$_{\rm ellip}= 4.0 \cdot 10^{-3}$ag and   $\Delta$V$_{\rm ref}=4.3\cdot 10^{-3}$ mag (Morris \& Naftilan \cite{Morris93}).\\
\indent 
{\bf V1149 Ori (HD\,37824)} is an SB1 \object{RS CVn}-type variable with a primary K0 giant (Fekel \& Henry \cite{Fekel05}) and colors from UBVRI photometry consistent with a K2/3 III + F8 V binary system (Cutispoto, Messina \& Rodon\`o \cite{Cutispoto01}).  V1149 Ori has an orbital period of   P$_{\rm orb} = 53.57465^d$ (Fekel et al. \cite{Fekel87}) and a variable rotational photometric period with a mean value of  P$_{\rm pho} =  53.12^d$ (Fekel \& Henry \cite{Fekel05}). If granted the spectral classification by Cutispoto (\cite{Cutispoto95}), the observed photometric variability can be completely attributed to the giant component. The non-active earlier-type component can give a  significant flux contribution to the U band where the total fluxes from both components tend to be comparable, as far as the magnetic activity makes the late-type component fainter. We adopt for the hot and cool components the values of parameters proper for their spectral classes: T$_h = 6200$\,K, T$_c = 4600$\,K, R$_h = 1.15$ R$_{\odot}$, R$_c = 12.6$ R$_{\odot}$, M$_h$/M$_c$ $\simeq$ 0.90,  and log\,$g_h$ = 4.5, log\,$g_c$ = 2.0 cm s$^{-2}$ (Cox 2000) and an assumed value of $i$ = 60$^{\circ}$. We find that the variability of V1149 Ori arising from proximity and reflection effects is about $\Delta$V$_{\rm ellip}= 1.3 \cdot 10^{-3}$ mag and   $\Delta$V$_{\rm ref}=6.3\cdot 10^{-4}$ mag (Morris \& Naftilan \cite{Morris93}).  \\
\indent 
{\bf \object{DH Leo} (HD\,86590)} is a triple system consisting of a BY-Dra type K0\,V + K7\,V binary with an orbital period of P$_{\rm orb} \sim 1.070354^d$ (Bolton et al. \cite{Bolton81}) and a K5\,V tertiary component (Barden \cite{Barden84}). The observed photometric variability is dominated by the K0\,V component, whose U, B and V total fluxes are larger than the fluxes from the K5\,V and K7\,V. We adopt for the components the values of parameters proper for their spectral classes:  T$_h = 5300$\,K, T$_c = 4300$\,K, T$_3 = 4600$\,K, R$_h = 0.81$ R$_{\odot}$, R$_c = 0.65$ R$_{\odot}$, R$_3 = 0.68$ R$_{\odot}$, M$_{\rm K0V}$/M$_{K7V}$ $\simeq$ 1.3, $i$ = 80$^{\circ}$ (Strassmeier et al. \cite{Strassmeier93}), log\,$g_h$ = 4.5, log\,$g_c$ = 4.5, log\,$g_3$ = 4.5 cm s$^{-2}$ (Cox \cite{Cox00}).  We find that the variability of \object{DH Leo} arising from proximity and reflection effects is about $\Delta$V$_{\rm ellip}= 1.0 \cdot 10^{-3}$ mag and   $\Delta$V$_{\rm ref}=2.8\cdot 10^{-3}$ mag (Morris \& Naftilan \cite{Morris93}). \\
\indent 
{\bf \object{HU Vir} (HD\,106225)} is an SB1 \object{RS CVn}-type  system with a K1 subgiant (Cutispoto \cite{Cutispoto98}) and an orbital period of  P$_{\rm orb} = 10.38758^d$ (Strassmeier \cite{Strassmeier94}). The rotational photometric period of about P$_{\rm pho} = 10.41^d$ is variable (Strassmeier et al. \cite{Strassmeier97}). In our model we adopt as effective temperature, radius and surface gravity the values proper for the  subgiant component:  T$_c = 5000$\,K, R$_c = 8$ R$_{\odot}$, log\,$g_c$ = 3.5 cm s$^{-2}$ (Cox \cite{Cox00}). \object{HU Vir} will be treated in the analysis as a single star and proximity and reflection effects will be considered negligible.\\
\indent 
{\bf \object{RS CVn} (HD\,114519)} is an eclipsing detached binary system consisting of a K2 subgiant and a F5 main-sequence star (Reglero et al. \cite{Reglero90}). This system has a mean orbital period of P$_{\rm orb} = 4.797855^d$ (Catalano \& Rodon\`o \cite{Catalano74}). The observed photometric variability can be entirely attributed to the subgiant, which is the only magnetically active component. However, the earlier-type component has a slightly larger total flux in the U, B and V bands, which can give a significant contribution to the observed blueing  as  the brightness of the active component decreases (Aarum Ulvas \& Henry \cite{Aarum05}). We adopt for the hot and cool component the values of parameters from Rodon\`o et al. (\cite{Rodono95}; \cite{Rodono01a}): T$_h = 6300$\,K, T$_c = 4600$\,K, R$_h = 1.89$ R$_{\odot}$, R$_c = 3.85$ R$_{\odot}$, M$_h$/M$_c$ = 0.96, $i$ = 85$^{\circ}$, and log\,$g_h$ = 4.5, log\,$g_c$ = 3.5 cm s$^{-2}$.  We find that the variability of \object{RS CVn} arising from proximity and reflection effects is about $\Delta$V$_{\rm ellip}= 0.01$ mag and   $\Delta$V$_{\rm ref}=4.7\cdot 10^{-3}$ mag (Morris \& Naftilan \cite{Morris93}).\\
\indent 
{\bf \object{V775 Her} (HD\,175742)} is an SB1 BY-Dra type variable consisting of a K0 and a K5/M2 main-sequence stars and with an orbital period of P$_{\rm orb} = 2.879395^d$ (Imbert \cite{Imbert79}). Although both components are expected to be magnetically active, due to their short rotation period and late-spectral type, the luminosity of the K0 component largely dominates the system's luminosity. We adopt for the hot and cool components values of parameters proper for their spectral classes: T$_h = 5300$\,K, T$_c = 4000$\,K, R$_h = 0.85$ R$_{\odot}$, R$_c = 0.60$ R$_{\odot}$, M$_h$/M$_c$ $\simeq$ 1.3, and log\,$g_h$ = 4.5, log\,$g_c$ = 4.5 cm s$^{-2}$ (Cox 2000)  and an assumed value of $i$ = 60$^{\circ}$.  We find that the variability of \object{V775 Her} arising from proximity and reflection effects is about $\Delta$V$_{\rm ellip}= 1.7\cdot 10^{-4}$ mag and   $\Delta$V$_{\rm ref}=6.2\cdot 10^{-4}$ mag (Morris \& Naftilan \cite{Morris93}).\\
\begin{figure*}
\begin{minipage}{18cm}
\centerline{
\psfig{file=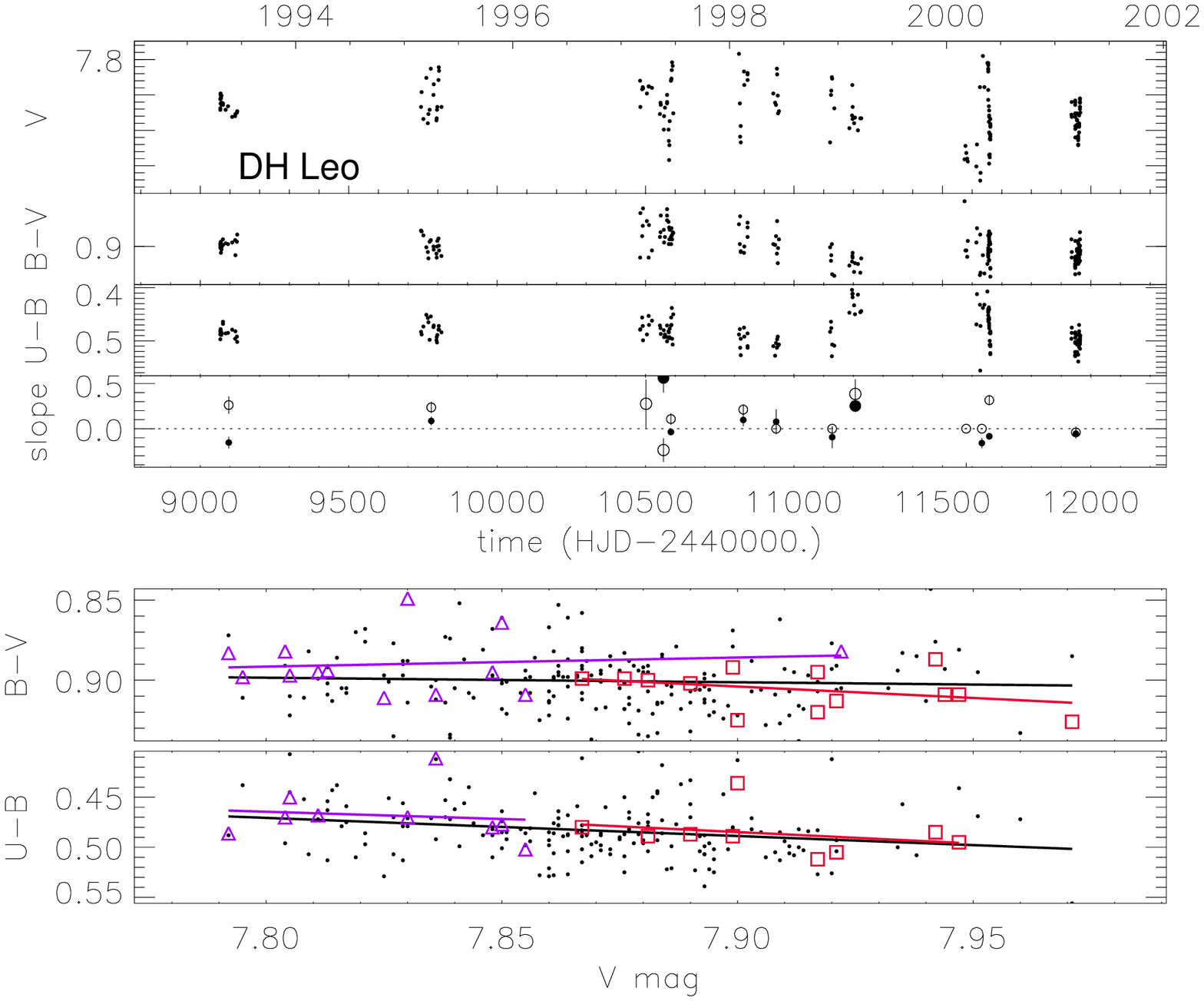,width=10cm,height=10cm,angle=90}
\psfig{file=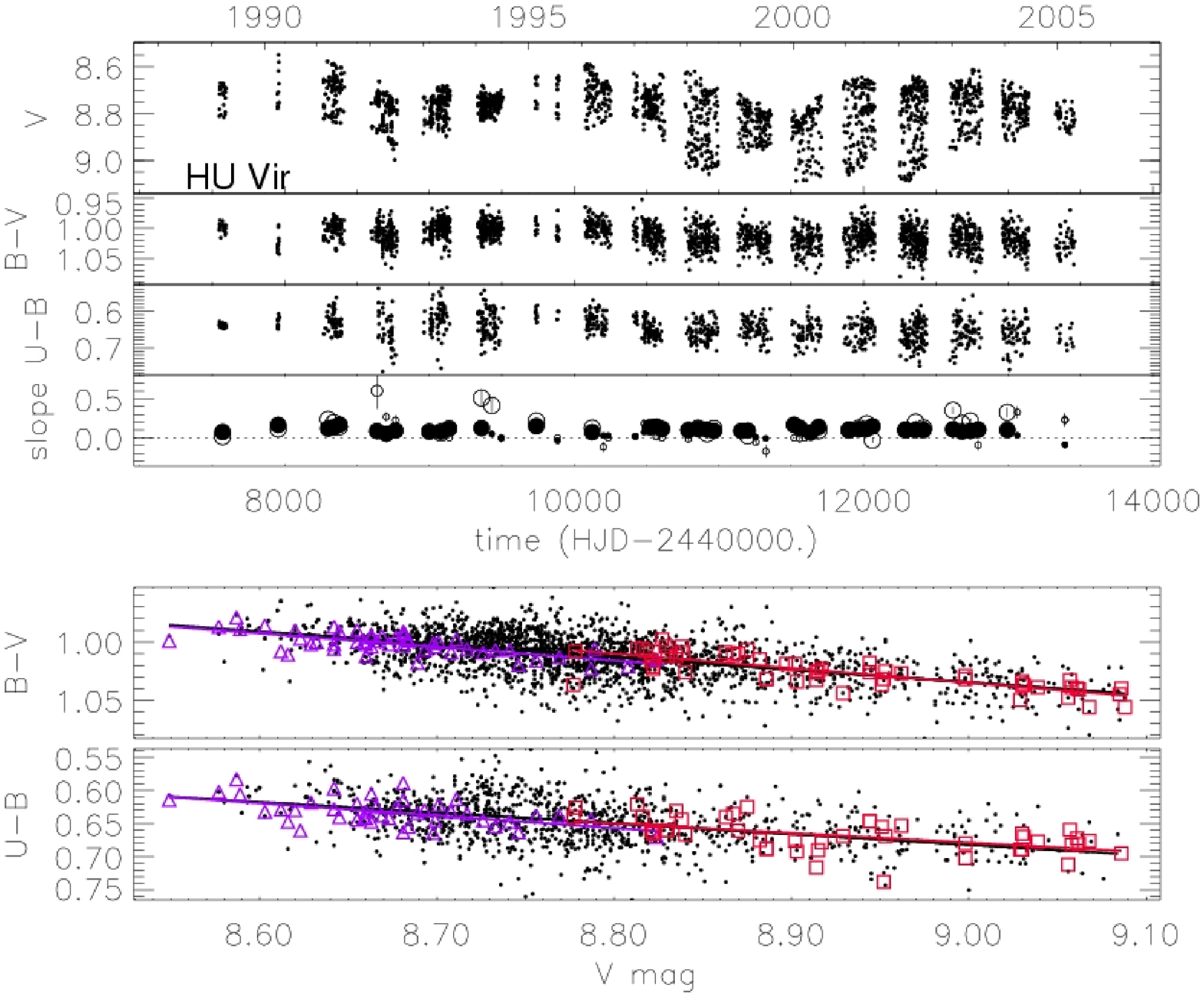,width=10cm,height=10cm,angle=90}
}
\end{minipage}
\caption{As in Fig.~\ref{fig-arpsc}, but for \object{DH Leo} and \object{HU Vir}. }
\end{figure*}
\indent 
{\bf \object{AR Lac} (HD\,210334)} is an eclipsing detached binary system of RS-CVn type consisting of a G2 and a K2 subgiants (Hall \cite{Hall76}). \object{AR Lac} has an orbital period of about  P$_{\rm orb} = 1.9832142^d$ (Jetsu et al. \cite{Jetsu97}). The observed photometric variability can be mostly attributed to the K2 subgiant, due to the deeper convection zone with respect to the G2 component. Nonetheless, since both components have similar total fluxes in U, B and V bands, the hotter component may play some role in the observed color variations, as it will be discussed in the following. We adopt for the hot and cool components the values of parameters from Lanza et al. (\cite{Lanza98}): T$_h = 5560$\,K, T$_c = 4820$\,K, R$_h = 1.51$ R$_{\odot}$, R$_c = 2.61$ R$_{\odot}$, M$_h$/M$_c$ = 0.97, $i$ = 87$^{\circ}$, and log\,$g_h$ = 4.0, log\,$g_c$ = 3.5 cm s$^{-2}$. We find that the variability of \object{AR Lac} arising from proximity and reflection effects is about $\Delta$V$_{\rm ellip}= 0.027$ mag and   $\Delta$V$_{\rm ref}=0.010$ mag (Morris \& Naftilan \cite{Morris93}).\\
\indent 
{\bf \object{SZ Psc} (HD\,219113)} is an eclipsing detached binary system belonging to the \object{RS CVn} class of variable stars and consisting of an F8 V/IV and a K1 subgiant (Hall \cite{Hall76}).
This system has an orbital period of P$_{\rm orb} = 3.9657889^d$. The observed photometric variability arises from the subgiant component which is the only magnetically active component. However, the earlier-type component can give a significant flux contribution to the U band as far as the magnetic activity makes the K1 component fainter. We adopt for the hot and cool components the values of parameters from Lanza et al. (\cite{Lanza01}) and Eaton \& Henry (\cite{Eaton07}): T$_h = 6100$\,K, T$_c = 4900$\,K, R$_h = 1.5$ R$_{\odot}$, R$_c = 6.0$ R$_{\odot}$, M$_h$/M$_c$ = 0.76, $i$ = 69$^{\circ}$, and log\,$g_h$ = 4.20, log\,$g_c$ = 3.23 cm s$^{-2}$. We find that the variability of \object{SZ Psc} arising from proximity and reflection effects is about $\Delta$V$_{\rm ellip}= 0.012$ mag and   $\Delta$V$_{\rm ref}=3.3\cdot 10^{-3}$ mag  (Morris \& Naftilan \cite{Morris93}).\\
\indent 
{\bf \object{II Peg} (HD\,224085)} is an SB1 \object{RS CVn}-type binary system with a primary K2\,IV component (Rucinski \cite{Rucinski77}). It has a variable rotational photometric period of about P$_{\rm pho} = 6.720^d$ (Rodon\`o et al. \cite{Rodono00}). We adopt as effective temperature the values from Rodon\`o et al. (\cite{Rodono00}), and radius and surface gravity values proper for its spectral class: T$_h = 4600$\,K, R$_h = 8$ R$_{\odot}$, log\,$g_h$ = 3.50 cm s$^{-2}$. \object{II Peg} will be treated in the analysis as a single star and proximity and reflection effects will be considered negligible.\\
\indent 
{\bf \object{BY Dra} (HD\,234677)} is an SB2 \object{BY Dra}-type binary system consisting of two M0 main-sequence stars (Rodon\`o \& Cutispoto \cite{Rodono92}). This system has an orbital period of P$_{\rm orb} = 5.976^d$ (Bopp \& Evans \cite{Bopp73}) which significantly differs from the rotational photometric period of about P$_{\rm pho} = 3.836^d$ (Rodon\`o et al. \cite{Rodono83}). Both components are expected to equally contribute to the observed photometric variability, their rotation rate and spectral type being equal. We adopt for the components the values of parameters proper for their spectral classes: T$_h = 3700$\,K, R$_h = 0.60$ R$_{\odot}$, M$_h$/M$_c$ = 1.0, and log\,$g_h$ = 4.50 cm s$^{-2}$ (Cox \cite{Cox00})  and an assumed value of $i$ = 60$^{\circ}$. We find that the variability of \object{BY Dra} arising from proximity and reflection effects is about $\Delta$V$_{\rm ellip}= 5.5\cdot 10^{-5}$ mag  and   $\Delta$V$_{\rm ref}=4.0\cdot 10^{-4}$ mag  (Morris \& Naftilan \cite{Morris93}).\\
\indent
Brightest V magnitude, maximum V-band light curve amplitude and average B$-$V and U$-$B colors of the program stars are listed in Table\,\ref{target}.

\begin{table}[!h]
\caption{\label{tab-comparison} V magnitude,  B$-$V and U$-$B colors of comparison (C) and check (CK) stars.}
\begin{tabular}{|r l| l@{\hspace{.1cm}} l| c@{\hspace{.1cm}} r@{\hspace{.1cm}} r|}
\hline
 \multicolumn{2}{|c|}{Program Star} & \multicolumn{2}{|c|}{C/CK Star} &V & B$-$V & U$-$B\\
 \multicolumn{2}{|c|}{}     &  \multicolumn{2}{|c|}{}   &  (mag)   &(mag)    & (mag)\\
\hline
1&\object{AR Psc}   & HD 7446   &(C)  & 6.04    & 1.08    & 1.02 \\
 &         & HD 7804   &(CK) & 5.16    & 0.07    & 0.03 \\
 2&\object{VY Ari}   & HD 17572  &(C)  & 6.72    & 0.37    & $-$0.03\\
 &         & HD 17329  &(CK1)& 7.93    & 0.76    & 0.28\\
  &        & HD 17395  &(CK2)& 8.45    & 0.59    & 0.06\\
 &         & HD 16187  &(CK3)& 6.05    & 1.05    & 0.88\\
3&\object{UX Ari}   & HD 20825  &(C)  & 5.55    & 1.10    & 0.92\\
 &         & HD 20618  &(CK1)& 5.91    & 0.86    & 0.45\\
 &         & HD 20644  &(CK2)& 4.47    & 1.55    & 1.86\\
4&\object{V711 Tau} & HD 22796  &(C)  & 5.55    & 0.94    & 0.67\\
 &         & HD 22819  &(CK1)& 6.10    & 0.99    & 0.78\\
 &	 & HD 23413  &(CK2)& 5.56    & 1.42    & 1.72\\
 &	 & HD 22484  &(CK3)& 4.29    & 0.57    & 0.05\\ 
5&\object{EI Eri}   & HD 25852  &(C)  & 7.83    & 1.01    & 0.78\\ 
  &        & HD 25954  &(CK1)& 7.53    & 1.24    & 1.36\\
  &        & HD 26237  &(CK2)& 7.17    & 0.03    & 0.03\\
  &        & HD 26409  &(CK3)& 5.44    & 0.94    & 0.67\\
6&V1149 Ori& HD 37741  &(C)  & 8.19    & 1.07    & 0.87\\
 &	 & HD 38145  &(CK1)& 7.91    & 0.33    & 0.06\\
 &	 & HD 38529  &(CK2)& 5.93    & 0.78    & 0.48\\
 &	 & HD 37984  &(CK3)& 4.91    & 1.17    & 1.10\\
7& \object{DH Leo}& HD 86132  & (C) & 8.17    & 0.96    & 0.65\\
 &       & HD 85428   & (CK1)& 7.78   & 1.24    & 1.33\\
 &       & HD 88008   &(CK2)& 8.48   & 0.74    & 0.26\\
 &       & HD 85376   & (CK3)& 5.30  & 0.22    & 0.09\\
8& \object{HU Vir}   & HD 106270 & (C) & 7.59    & 0.74    & 0.30\\
 &	 & HD 105796 &(CK1)& 8.05    & 1.07    & 0.96\\
 &	 & HD 105759 &(CK2)& 6.55    & 0.22    & 0.07\\
 &	 & HD 106516 &(CK3)& 6.12    & 0.46    & $-$0.14\\ 
9 &\object{RS CVn}   & HD 114778 &(C)  & 8.40    & 0.47    & $-$0.04\\
 &	 & HD 114838 &(CK1)& 8.08    & 0.53    & 0.01\\
 &	 & HD 114863 &(CK2)& 8.54    & 0.57    & 0.02\\
 &	 & BD+362347 &(CK3)& 9.91    & 0.50    &$-$0.09\\
 &	 & HD 113797 &(CK4)& 5.22    &$-$0.07  &$-$0.19\\
 &	 & HD 114357 &(CK5)& 6.02    & 1.17    & 1.16\\
10 &\object{V775 Her} & HD 337271 &(C)  & 8.74    & 1.12    & 1.00\\
  &        & HD 337275 &(CK1)& 8.55    & 1.25    & 1.30\\
 &	 & HD 178187 &(CK2)& 5.85    & 0.09    & 0.19\\
 &	 & HD 174160 &(CK3)& 6.22    & 0.48    & 0.02\\
 &	 & HD 175492 &(CK4)& 4.60    & 0.78    & 0.48\\ 
11 &\object{AR Lac}   & HD 208728 &(C)  & 6.81    & 1.19    & 0.85\\
 &	 & HD 209945 &(CK1)& 5.10    & 1.58    &1.95\\
 &	 & HD 209219 &(CK2)& 7.40    & 1.35    &1.56\\
 &	 & HD 210731 &(CK3)& 7.45    & 0.57    &0.04\\
12 & \object{SZ Psc}& HD 219018 & (C) & 7.74    & 0.64    & 0.15\\
 &       & HD 219150  & (CK1)& 7.20   & 0.40    & $-$0.07\\
 &       & HD 218527  & (CK2) & 5.46 & 0.91 & 0.56\\
 13 &\object{II Peg}   & HD 224016 &(C)  & 8.52    & 0.78    & 0.43\\
  &        & HD 223461 &(CK1)& 5.96    & 0.20    & 0.05\\
 &	 & HD 224895 &(CK2)& 6.81    & 1.21    & 1.17\\
14 &\object{BY Dra}   & HD 172268 &(CK1)& 7.89    & 1.24    & 1.31\\
  &        & HD 172468 &(CK2)& 7.52    & 1.25    & 1.16\\
 &	 & HD 169028 &(CK3)& 6.29    & 1.10    & 1.08\\
 &	 & HD 172883 &(CK4)& 5.98   &$-$0.07   &$-$0.23\\
\hline  
\end{tabular}
\end{table}

\begin{table}
\caption{Total number of averaged observations in the V, B and U band, total number of light curves and interval of time of the photometric monitoring \label{tab-obs}}
\begin{tabular}{lrrrrr}
\hline
Program Star &  V  & B   & U  & L.C.  & Time Range \\
\hline
\object{AR Psc} & 620  & 595 & 555 & 35 & 1987-2004 \\
\object{VY Ari} & 980  & 966 & 904 & 40 & 1991-2004 \\
\object{UX Ari} & 1169 & 1157& 1088& 37 & 1990-2004 \\
\object{V711 Tau}& 747 & 733 & 669 & 31 & 1990-2004 \\
\object{EI Eri} & 961 & 934 & 829 & 43 & 1987-2005  \\
V1149 Ori & 1247 & 1226 & 1114 & 35 & 1989-2004 \\
\object{DH Leo} & 184 & 184 & 179 & 13 & 1993-2001 \\
\object{HU Vir} & 2182 & 2021 & 1006 & 52 & 1989-2004 \\
\object{RS CVn} & 1900 & 1019 & 969 & 19 & 1990-2004 \\
\object{V775 Her} & 1105 & 1048 & 564 & 54 & 1990-2004 \\
\object{AR Lac} & 2348 & 698 & 620 & 23 & 1990-2004 \\
\object{SZ Psc} & 185 & 84 & 80 & 8 & 1993-1998 \\
\object{II Peg} & 1233 & 1193 & 1049 & 43 & 1992-2004 \\
\object{BY Dra} & 734 & 725 & 391 & 42 & 1990-2004 \\

\hline
\end{tabular}
\end{table}
\section{Observations}

The photometric observations were collected by two APTs:
the \it Phoenix-25 \rm since 1988, and the \it APT80/1 \rm since late 1992. The  Phoenix-25 is a 25-cm telescope operated under the ''rent-a-star'' service  by the Franklin \& Marshall College at Washington Camp (AZ, USA). It
feeds a single-channel photon counting photometer, equipped with 
an uncooled 1P21 photomultiplier and standard {\sl UBV} filters (Boyd et al. \cite{Boyd84}; Baliunas et al. \cite{Baliunas85}). 
The APT80/1 is a 80-cm telescope located at the {\it M.~G.~Fracastoro} station of OAC on Mt.~Etna
(1725m a.s.l.) that feeds a single channel charge-integration photometer, equipped with an uncooled Hamamatsu R1414 SbCs 
photomultiplier and standard {\sl UBV} filters (Rodon\`o \& Cutispoto \cite{Rodono92}; Messina \cite{Messina98}).\\
\indent
The Phoenix-25 observed the program stars differentially with respect to the comparison (C) and check (CK1) stars listed 
in Table \ref{tab-comparison}. A detailed description of the observation and reduction procedures for the data collected with this telescope 
can be found in Rodon\`o \& Cutispoto (\cite{Rodono92}).\\
\indent
The APT80/1 observed all the program stars differentially with respect to a larger set of comparison stars (all stars in Table \ref{tab-comparison}) including those observed by the Phoenix-25. The integration time in U, B and V filters was set to 15, 10 and 10  s, respectively,  and the observing sequence was {\sl n-c-ck1-c-v-v-v-c-v-v-v-c-ck2-c-n}, where the symbol $n$ denotes the bright navigation star, which is the first star of the group the APT80/1 hunts. The sky background was measured  at a fixed position near each star.\\
\indent
Differential magnitudes from both telescopes were corrected for atmospheric extinction  and transformed into the standard Johnson UBV system. The transformation coefficients were determined quarterly by observing  selected samples of standard stars. Due to the relatively short duration of an observing sequence ($\simeq 30$ minutes), differential values were finally averaged to obtain one single average point. The complete dataset presented in this paper consists of about, 16,000, 12,600 and 10,000 average points in the V, B and U filters (see Table\,\ref{tab-obs}).
After transformation into the standard system the achieved accuracy of V magnitude, B$-$V and U$-$B color indices is of the order of 0.008, 0.010 and 0.012 mag, respectively, for the faintest stars (V$\simeq$8.5 mag).
A comparison between the  standard deviations of ck$-$c  and v$-$c differential magnitudes shows that the comparison stars have remained constant within the observation accuracy.\\
\indent
We could homogenize very well the observations coming from two different telescopes, since they observed a common set of comparison and check stars for many years. \\
\indent
In order to further extend our time series, in the following analysis we made use also of the UBV observations  of \object{AR Psc}, \object{EI Eri}, V1149 Ori and \object{HU Vir} collected  with the 50-cm telescope at ESO (La Silla, Chile) published by Cutispoto (\cite{Cutispoto90}; \cite{Cutispoto92}; \cite{Cutispoto93}).
For the eclipsing binary stars \object{AR Lac}, \object{RS CVn} and \object{SZ Psc},  we will consider only the out-of-eclipse observations, being our analysis focused only on the  magnitude and color variations arising from the presence of photospheric inhomogeneities. In order to determine the out-of-eclipse phases, we also took into account the orbital period variations.
\begin{figure*}
\begin{minipage}{18cm}
\centerline{
\psfig{file=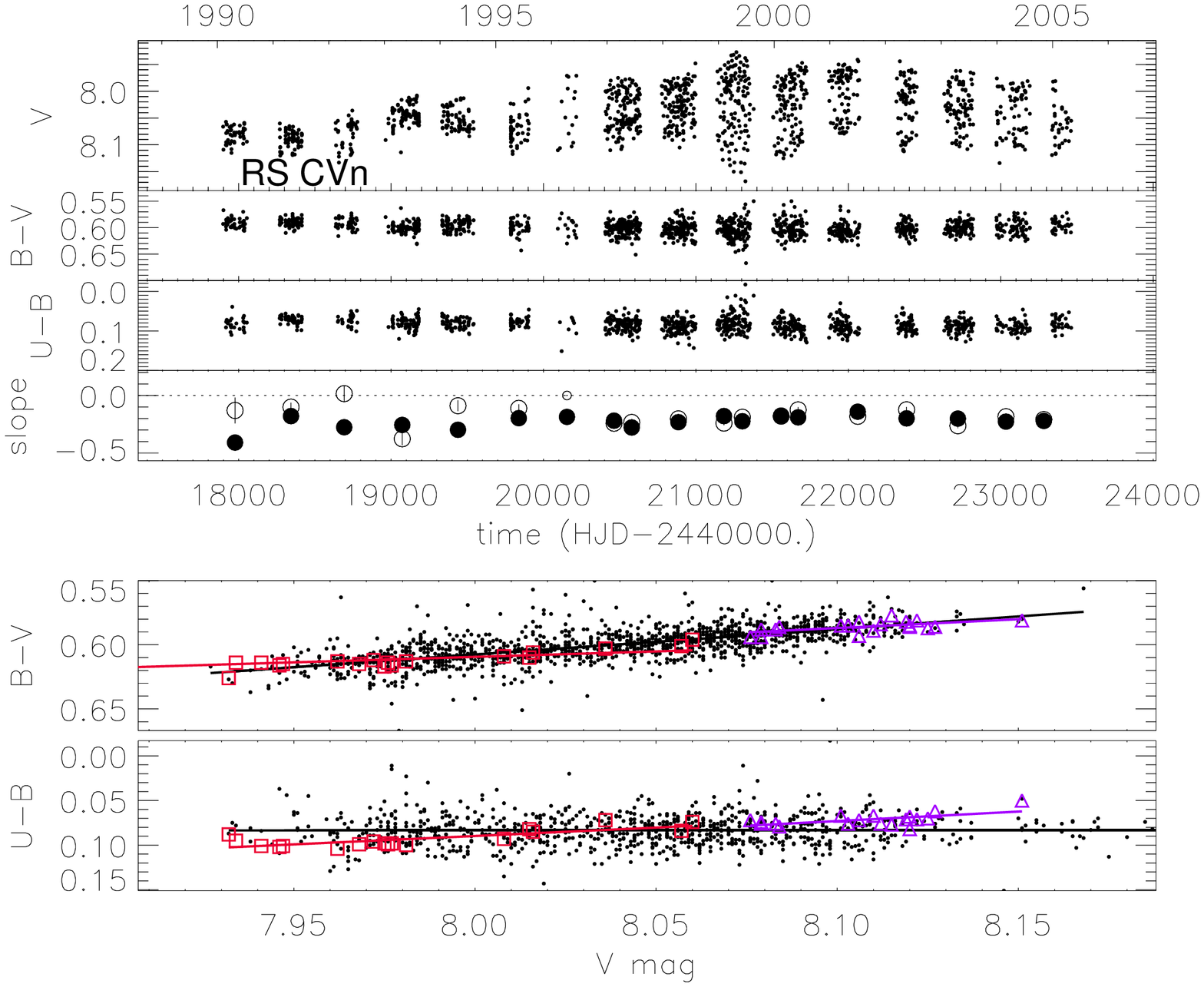,width=10cm,height=10cm,angle=90}
\psfig{file=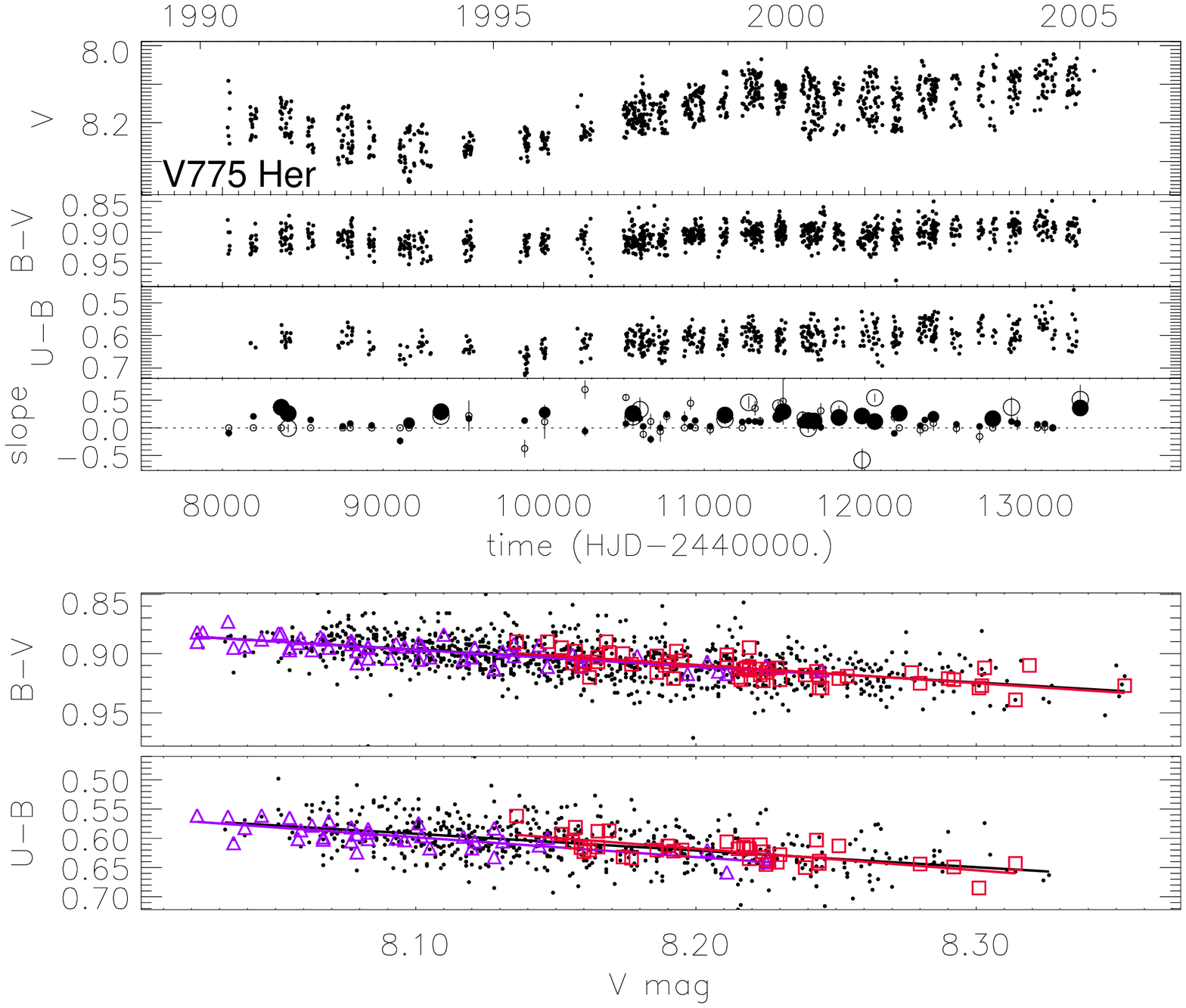,width=10cm,height=10cm,angle=90}
}
\end{minipage}
\caption{ As in Fig.~\ref{fig-arpsc}, but for \object{RS CVn} and \object{V775 Her}. }
\end{figure*}
\begin{figure*}
\begin{minipage}{18cm}
\centerline{
\psfig{file=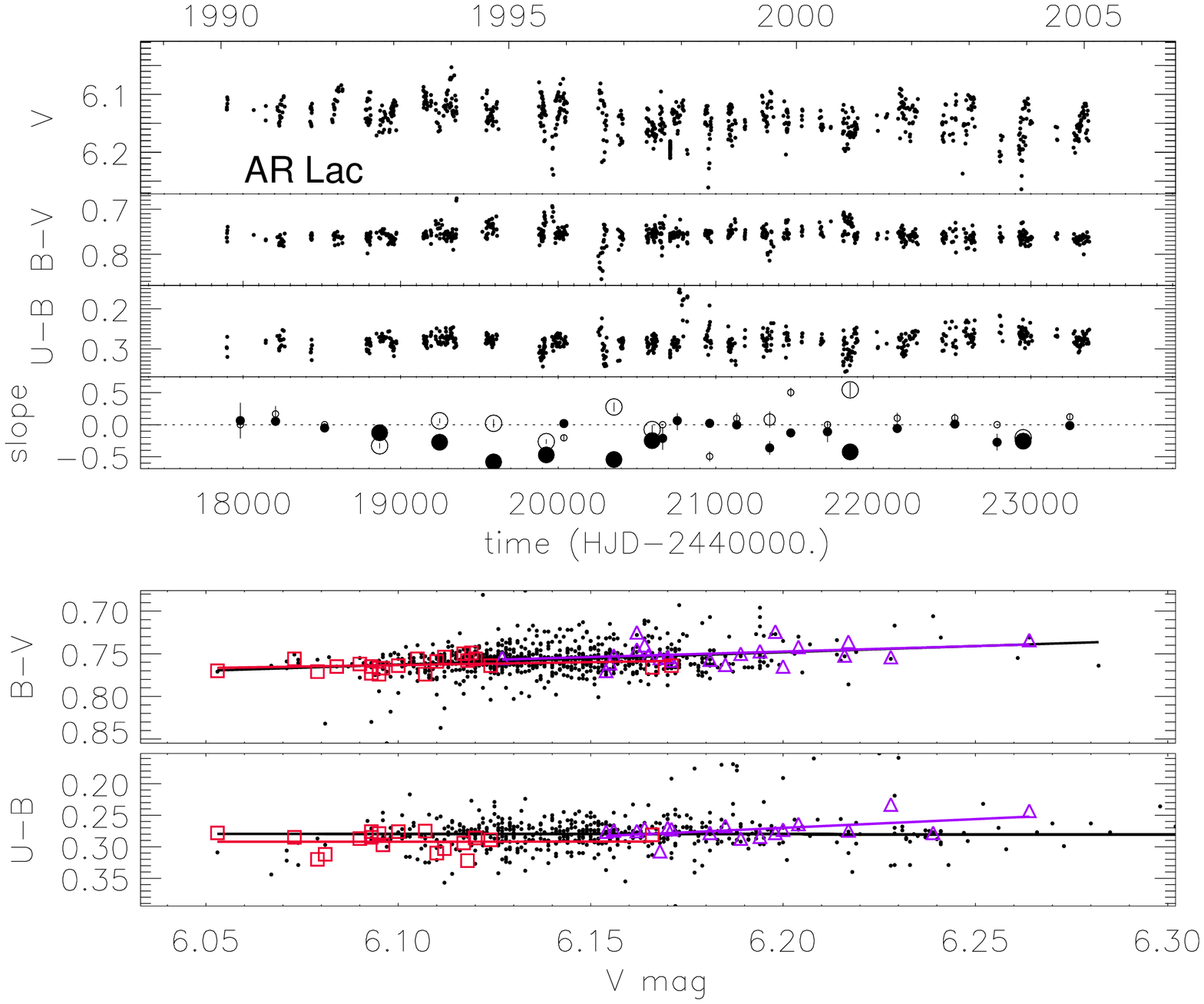,width=10cm,height=10cm,angle=90}
\psfig{file=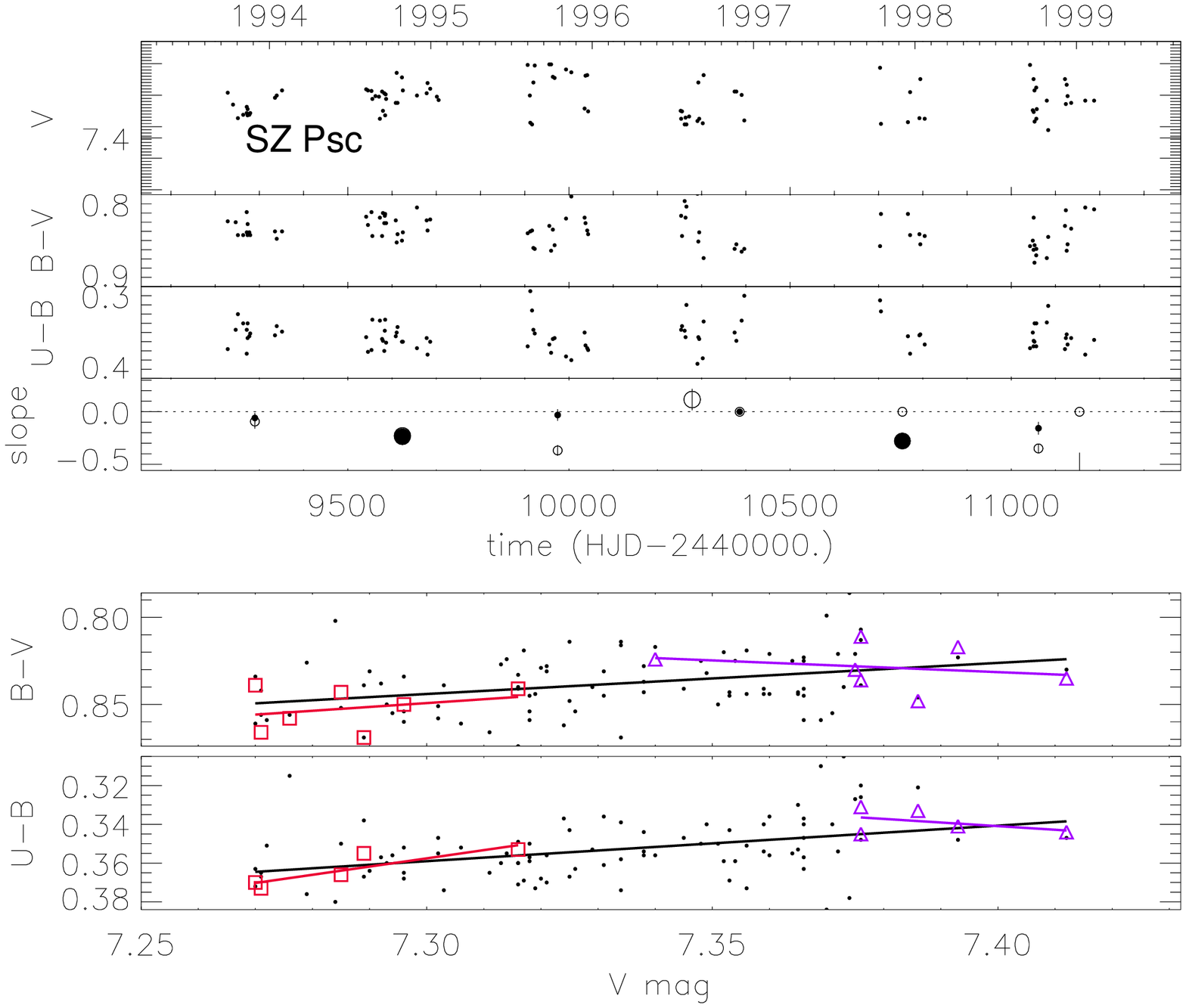,width=10cm,height=10cm,angle=90}
}
\end{minipage}
\caption{As in Fig.~\ref{fig-arpsc}, but for \object{AR Lac} and \object{SZ Psc}. }
\end{figure*}

\begin{figure*}
\begin{minipage}{18cm}
\centerline{
\psfig{file=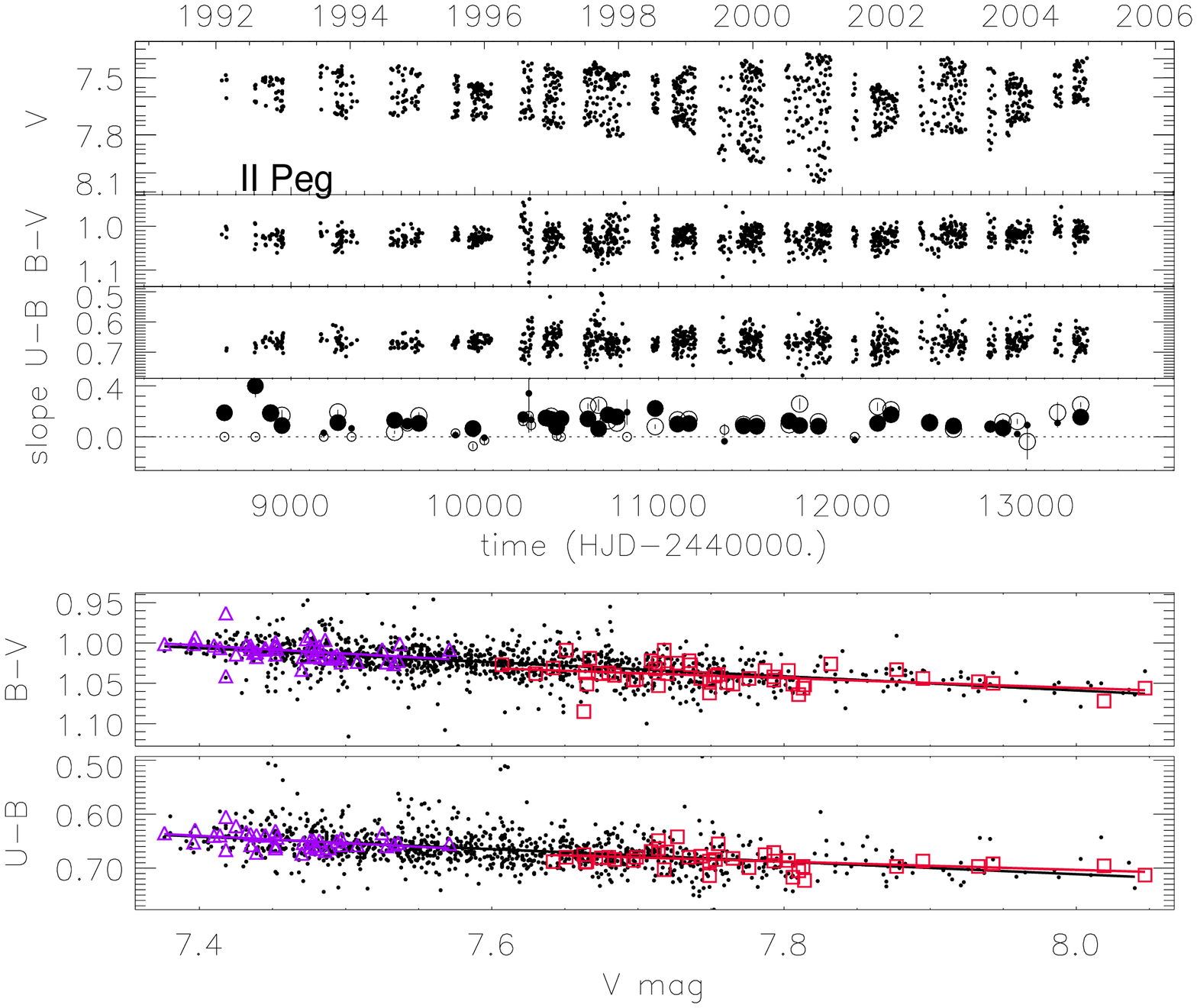,width=10cm,height=10cm,angle=90}
\psfig{file=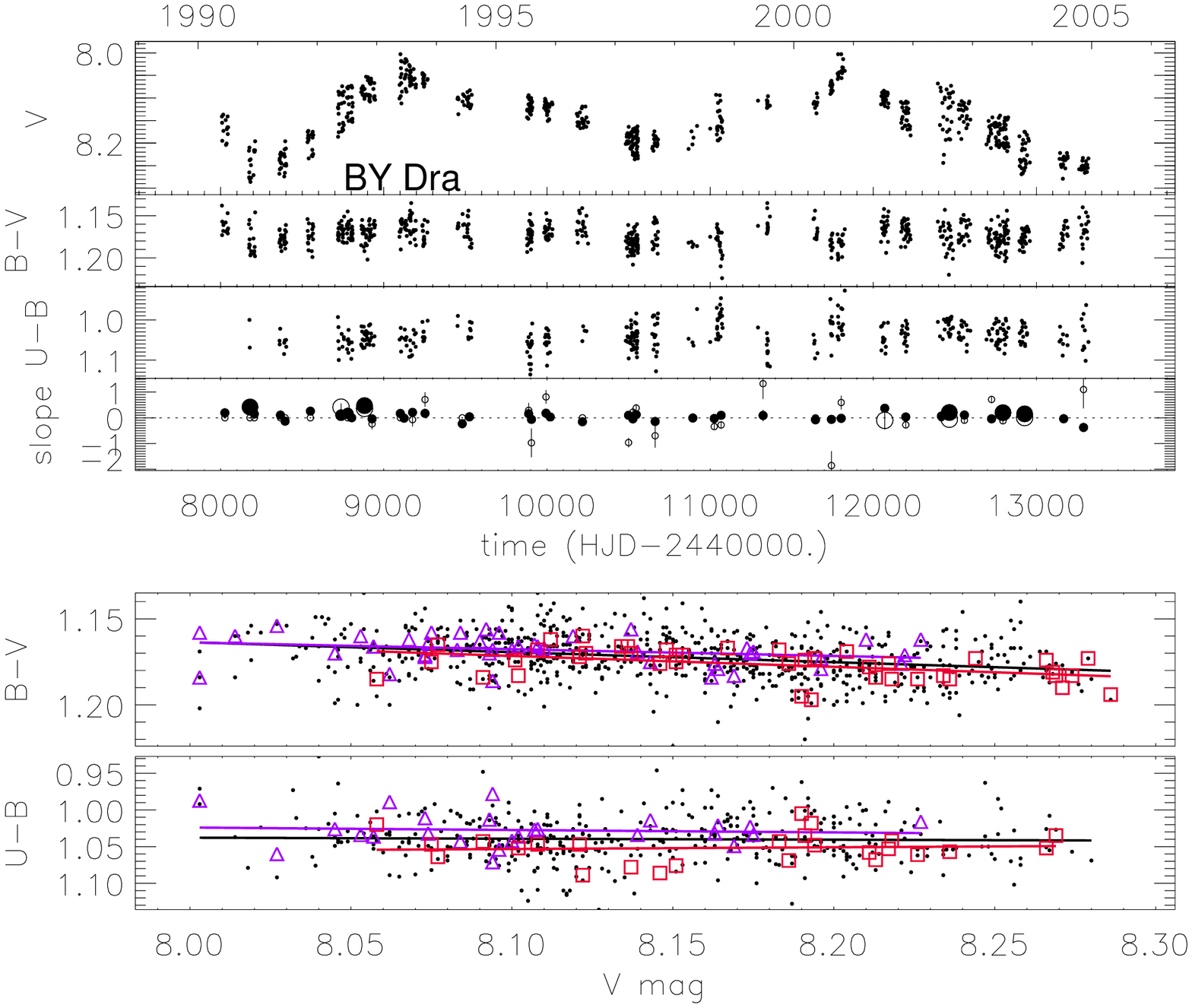,width=10cm,height=10cm,angle=90}
}
\end{minipage}
\caption{\label{fig-bydra} As in Fig.~\ref{fig-arpsc}, but for \object{II Peg} and \object{BY Dra}. }
\end{figure*}
\indent
In order to better investigate the evolution of shape, amplitude and mean magnitude shown by the light curves, the whole data set of each program star was subdivided into a number of light curves. The division was made by selecting time intervals
during which the star displayed a stable flux modulation, i.e. no significant differences (smaller than $\sim$ 0.01-0.02 mag) between observations falling close to each other within 0.01-0.02 dex in rotational phase. That division allowed us to obtain from a minimum of 8 light curves in the case of \object{SZ Psc} to a maximum of 54 light curves in the case of \object{V775 Her}. In Fig.\,\ref{uxari_curva} we plot, as an example, one of the 37 light curves in which the complete series of observations of \object{UX Ari} has been divided. V magnitude, B$-$V and U$-$B colors for the mean epoch 1993.82 are plotted vs. rotational phase in the top three panels of Fig.\,\ref{uxari_curva}.
The most relevant properties of the light curves of each program star are listed in the on-line Table 4. Specifically, we give: mean epoch of observations (Mean Epoch), mean (HJD$_{\rm mean}$), initial (HJD$_{\rm ini}$) and final  (HJD$_{\rm end}$) heliocentric Julian day, number of points in the light curve (N$_{\rm m}$), brightest V magnitude (V$_{\rm min}$) and light curve peak-to-peak amplitude in the V band ($\Delta$V), B$-$V and U$-$B colors, standard deviation ($\sigma_v$) of v$-$c and ($\sigma_{ck1-c}$) of ck1$-$c differential observations, and the telescope used to make the observations.\\
\indent
With the exception of \object{SZ Psc} and \object{DH Leo}, all the targets were observed in many epochs almost contemporaneously  by the two mentioned telescopes. Thanks to the difference in longitude between the two observation sites, that allowed to observe the same stars almost continuously for nine additional hours, we were able to obtain for the short-period stars a set of light curves well covered in rotational phase.\\

\section{V magnitude versus colors variations}

In Figs.\,\ref{fig-arpsc}-\ref{fig-bydra} we plot the complete series of V magnitudes (top panel), B$-$V and U$-$B colors (second and 
third panel from top) of our program stars. To date, this UBV database is among the longest and most homogeneous for the program close binary systems. Although their long- and short-term photometric behaviours have been investigated 
in a number of papers, differently than in previous works mostly based only on V-band data, our data allow us to investigate also the color behaviour on both the short and long timescales and by using a very homogeneous data sample.

\subsection{Short-term (rotational) variation}
\begin{figure}
\begin{minipage}{11cm}
\psfig{file=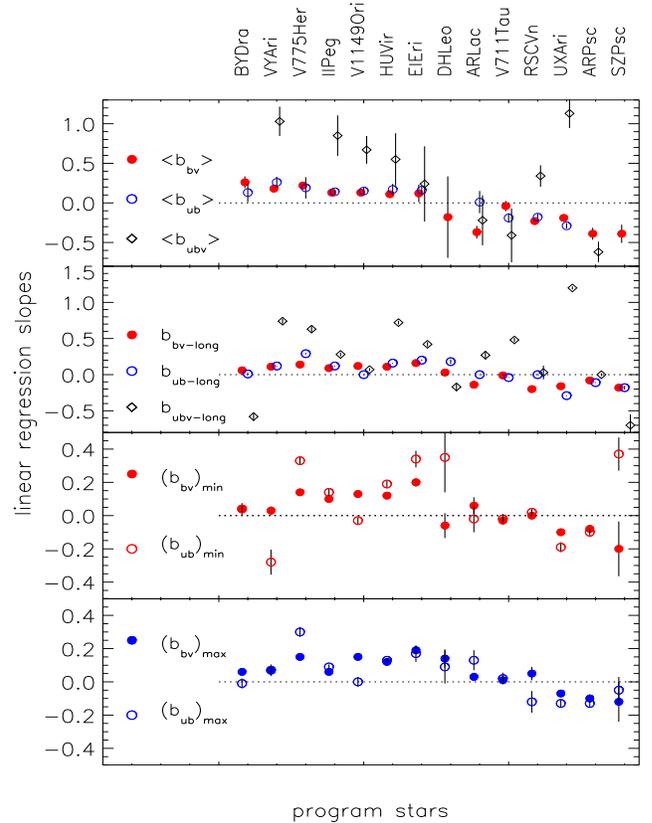,width=9.5cm,height=13cm}
\end{minipage}
\vspace{-1.5cm}
\caption{\label{fig-slopes} Average slopes of the color-magnitude and color-color relations ($\alpha > 0.1$) arising from rotational modulation ({\it top panel}) and from activity cycles or long-term trends ({\it second panel}). Slopes of the relations between brightest/bluest ({\it third panel})
and  between faintest/reddest light curves values ({\it fourth panel}).}
\end{figure}

The magnitude and color variations generally shown by active stars and related to magnetic activity have different time scales: the star's rotation modulates the visibility of asymmetrically distributed photospheric temperature inhomogeneities and, as a consequence, gives rise to magnitude and color variations with the same period of the star's rotation period. Within a few stellar rotations, the inhomogeneities grow and decay and change either amplitude and shape of the flux rotational modulation, as well as the mean brightness level. Finally, on a longer timescale,
variations of the inhomogeneities total area and their latitudinal migration, which are both related to the presence of starspot activity cycles on a differentially rotating star, cause additional magnitude and color variations (Messina, Rodon\`o \& Cutispoto \cite{Messina04}).
In order to separate the effects of rotation from those arising from ARGD and activity cycles, the complete time series of observations of each program star has been divided in a number of light curves, as mentioned in the previous section, i.e., each corresponding to an interval during which the star displayed a stable flux modulation. \\
\indent
For each light curve (see Fig.\,\ref{uxari_curva}) we have carried out correlation and regression analyses between colors and magnitude variations (B$-$V and U$-$B vs. V, and U$-$B vs. B$-$V), by computing correlation coefficients ($r$), their significance level ($\alpha$), and slopes of linear fits (see lower panels of  Fig.\,\ref{uxari_curva}). The significance level $\alpha$ represents the probability of observing a value of the correlation coefficient larger than $r$ for a random sample having same number of observations and degrees of freedom (Bevington \cite{Bevington69}). Correlation analysis allows us to investigate the origin of magnitude and color variations. For example, if these variations originate from a single spot or group of small spots, as well as if they originate from two different, but spatially and temporally correlated, types of inhomogeneities, e.g. cool spots and hot faculae, we expect these quantities to be correlated. 
On the contrary, a poor correlation or its absence will tell us that magnitude and colors are affected by at least two mechanisms, which are operating independently from each other, either spatially or temporally. As it will be better discussed in Sect.\,6, the presence of magnetic activity in the fainter stellar component of the system may also play some role in decreasing the expected correlation. The regression analysis is also important to infer relevant information on the properties of photospheric inhomogeneities, since their average temperature mostly determine the slope of the fits. Indeed, surface inhomogeneities with different areas but constant temperature, will determine magnitude and color variation of different amplitude but with a rather constant ratio. We have computed the slope $b$ of the linear fit $y=a+bx$ to B$-$V vs. V ($b_{bv}$), to U$-$B vs. V ($b_{ub}$) and to U$-$B vs. B$-$V ($b_{ubv}$) for each light curve;  $b_{bv}$ and $b_{ub}$ values are listed in the on-line Table\,5 and plotted in Figs.\,\ref{fig-arpsc}-\ref{fig-bydra} (fourth panel from top) with filled and open circles, respectively. The symbol size indicates different significance levels $\alpha$ of the correlation coefficients $r$: the largest size is for $\alpha \le 0.05$,  middle size for $0.05 < \alpha \le 0.1$, and smallest size for  $\alpha > 0.1$.
The values of the average slopes $<b_{bv}>$, $<b_{ub}>$ and $<b_{ubv}>$, uncertainty, number of light curves used to make the average (only curves with $\alpha < 0.1$) and the slopes minimum and maximum values are listed in the on-line Table\,6 and plotted in Fig.\,\ref{fig-slopes} (top panel)  with filled bullets, open bullets and diamonds, respectively.  Targets are just ordered with decreasing value of the average slope.\\
\indent
Correlation and regression analyses have given three important results: \\
\indent
\it i\,\rm)  magnitude and color variations are found to be correlated to each other only in a fraction of light curves. The percentage of light curves in which V magnitude and color variations are significantly  correlated ($\alpha < 0.1$) is over 60\% for  \object{VY Ari}, \object{II Peg}, V1149 Ori, \object{HU Vir}, \object{UX Ari} and \object{RS CVn}. These stars will be thereafter named \it color-correlated \rm stars. The percentage is smaller than 40\% for  \object{BY Dra}, \object{V775 Her}, \object{EI Eri}, \object{DH Leo}, \object{AR Lac}, \object{V711 Tau}, \object{AR Psc}, \object{SZ Psc}.  These stars will be thereafter named \it color-uncorrelated \rm stars. The B$-$V and U$-$B vs. V variations are generally found to be correlated more frequently than the U$-$B vs. B$-$V variations. \\
\indent
 \it ii\,\rm) the values of $<b_{bv}>$ and $<b_{ub}>$, as computed by considering values for which $\alpha < 0.1$, are positive for 8 program stars: \object{BY Dra}, \object{VY Ari}, \object{V775 Her}, \object{II Peg}, V1149 Ori, \object{HU Vir}, \object{EI Eri}, \object{DH Leo}. It means that along the star's rotation the fainter the star the redder its B$-$V and U$-$B colors. These stars will be thereafter named \it reddening \rm stars. Such a behaviour is consistent with a rotational flux modulation dominated by cool spots. 
The values of $<b_{bv}>$ and $<b_{ub}>$ are negative for 6 program stars: \object{AR Lac}, \object{V711 Tau}, \object{RS CVn}, \object{UX Ari}, \object{AR Psc}, \object{SZ Psc}. It means that along the star's rotation the fainter the star the bluer its B$-$V and U$-$B colors.  These stars will be thereafter named \it blueing \rm stars. Such a different behaviour may be consistent with a rotational flux modulation dominated either by cool spots, whose negative flux contribution dominates the V-band variation, and by bright faculae, whose positive flux contribution dominates the B- and U-band variations. However, a flux contribution to the B and U bands by an earlier-type stellar companion may also cause a similar blueing, as it will be discussed in Sect.\,5.\\
 \indent
 \it iii\,\rm) the values of the $b_{bv}$ and $b_{ub}$ slopes generally vary from light curve to light curve, often within a significant range of values (see the on-line Table\,5), indicating that the average temperature of surface inhomogeneities is variable.\\
 \indent
 Although \object{DH Leo} has a negative slope, however its associated uncertainty is large. As better shown in the following, it actually behaves like a  \it reddening \rm star.
Although the  $b_{ubv}$ value is expected to be positive for both \it reddening \rm and \it blueing \rm stars, it is found to be negative more frequently in \object{V711 Tau}, \object{AR Lac}, \object{AR Psc}, \object{BY Dra} (see on-line Table\,5). For instance, we find than $b_{ubv}$ is negative when the B$-$V and U$-$B vs. V variations are poorly correlated ($\alpha > 0.1$). In Sect.\,6 we will show that the presence of an active fainter companion or the presence a facular activity uncorrelated to the spots activity on the rotation time scale may be the reason for the negative slope of $b_{ubv}$.


\subsection{Long-term (cyclical) variation}
\begin{figure}
\begin{minipage}{11cm}
\psfig{file=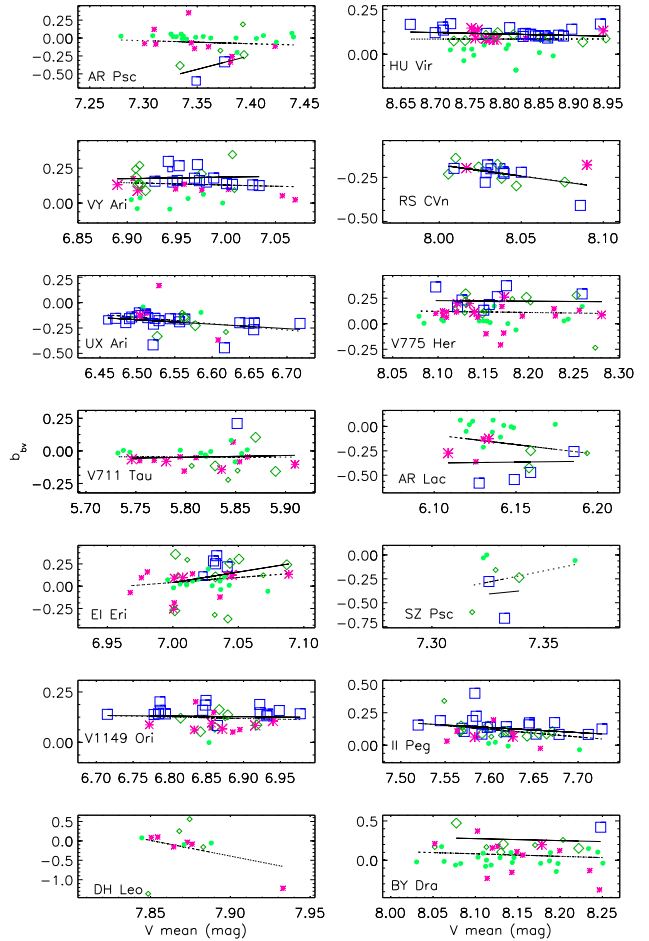,width=9cm}
\vspace{-1cm}
\end{minipage}
\caption{\label{slope_final} Slope b$_{bv}$ vs. mean V magnitude. Different symbols and sizes indicate different correlation coefficients and significance levels, respectively (see text). The solid and dotted lines are linear fits to significant values ($\alpha \le$ 0.1) and to all data, respectively. }
\end{figure}

In order to investigate the relation between color-magnitude and color-color variations arising from activity cycles or long-term trends we have made correlation and linear regression analyses to the complete series of data (small dots in the lower panels of Figs.\,\ref{fig-arpsc}-\ref{fig-bydra}). Correlation coefficients (r), significance levels ($\alpha$), values of the fit slope ($b_{bv-long}$, $b_{ub-long}$ and $b_{ubv-long}$), and number (N) of data used to compute the fits are listed in the on-line Table\,\ref{tabcoeff_long}.
The slopes $b_{bv-long}$, $b_{ub-long}$ and $b_{ubv-long}$ are plotted in the second panel of Fig.\,\ref{fig-slopes} as filled bullets, open bullets and diamonds, respectively. 
We have made separate fits also to the brightest and bluest values V$_{\rm min}$, (B$-$V)$_{\rm min}$, (U$-$B)$_{\rm min}$ of each light curve (open triangles in lower panels of Figs.\,\ref{fig-arpsc}-\ref{fig-bydra}) and to the faintest and reddest values V$_{\rm max}$, (B$-$V)$_{\rm max}$, (U$-$B)$_{\rm max}$ (open squares) and computed the slope values $(b_{bv})_{min}$, $(b_{ub})_{min}$ and $(b_{bv})_{max}$, $(b_{ub})_{max}$. These slopes are also listed in the on-line Table\,\ref{tabcoeff_long} and plotted in the third and fourth panel of Fig.\,\ref{fig-slopes}. Here we remind that the brightest magnitude and color values depend on those surface inhomogeneities evenly distributed along the stellar longitude; whereas, the faintest values depend on evenly plus unevenly distributed inhomogeneities.
We have also investigated the dependence of the slope values on the mean V magnitude, which is related to the phase of the starspot cycle. In a preliminary investigation, we have considered all the slope values, independently from their significance level, and  afterwards a subset of values with high significance ($\alpha \le 0.1$). The linear fits in both cases are generally in agreement, with the exclusion of \object{AR Psc} and \object{AR Lac} for which the number of significant values turned out to be small. The results of our investigation are plotted in Fig.\,\ref{slope_final}, where the symbol size indicates the significance of the correlation coefficient as in Fig.\,\ref{fig-arpsc}-\ref{fig-bydra}, whereas different symbols indicate different ranges of the correlation coefficients: filled circles for $0 < r \le 0.2$, asterisks for $0.2 < r \le 0.4$, diamonds  for $0.4 < r \le 0.6$, squares for $r \ge 0.6$. \\
\indent
Our regression analysis has given the following important results: \\
\indent
\it i\,\rm) our program stars show on  the long timescales the same behaviour found in the rotation timescale: when activity cycles and/or long-term trends make the star's brightness fainter, all \it reddening \rm stars become redder, whereas all \it blueing \rm stars become bluer. Exceptions are \object{V711 Tau}, whose long-term U$-$B and B$-$V color variations are found to be un-correlated to the V mag variations and  \object{AR Lac}, \object{RS CVn} and V1149 Ori whose U$-$B long-term color variations are scarcely correlated to the V mag variations. For instance, in the case of V1149 Ori only data from 2004/2005 make the value of correlation coefficient close to zero.\\
\indent
\it ii\,\rm)  the reddening slopes related to brightest values and arising from evenly distributed inhomogeneities are similar to the reddening slopes related to the faintest values and arising from evenly plus unevenly distributed inhomogeneities. An exception is represented by \object{VY Ari}, whose brightest/bluest slope is negative, instead of being positive. For instance, for the \it blueing \rm stars the brightest light curve values are found to be correlated to the reddest values and the faintest light curve values to the bluest. In other words, as the brightest light curve magnitude faints, the star gets bluer, and when the most spotted hemisphere is in view the star gets even bluer. 
We note that in the case of \object{BY Dra} and \object{SZ Psc} the color-color long-term variations, differently than expected, are anti-correlated. \\
\indent
\it iii\,\rm) the values of the $b_{bv}$ slope of \it reddening \rm stars is  rather independent or slightly decreasing at increasing values of the mean V magnitude.  The values of the $b_{bv}$ slope of \it blueing \rm stars becomes more negative (i.e. due to larger color variations) at increasing values of the mean V magnitude.
 We remind that very cool inhomogeneities do not produce color variation; on the contrary, the warmer the inhomogeneities the larger their color variations. Therefore, when our program stars approach the maximum level of starspot activity, the average temperature of inhomogeneities seems to be constant or slightly decreasing in \it reddening \rm stars, whereas it seems to be  increasing in \it blueing \rm stars. Such an increase may arise from an increased flux contribution by hot faculae, as well as by the earlier-type component.
Only the \it blueing \rm star \object{SZ Psc} and the \it reddening \rm star \object{EI Eri} deviate from this behaviour (see Fig.\,\ref{slope_final}).
On the other hand, a color variation may arise from the difference of limb-darkening between the most and the least spotted stellar hemisphere, depending on the average latitude in which spots are located.

\section{A simple modelling approach}
The analysis  so far carried out is not enough to understand whether the behaviour shown by \it blueing \rm stars arises from an enhanced flux contribution to the B and U bands by hot faculae or by the presence
of the earlier-type component in the binary system, whose relative flux contribution becomes larger when the active late-type component is made fainter by magnetic activity. Additional information on the possible origin of the observed  \it blueing \rm is here derived by adopting a first-order modelling. An accurate modelling of the light and color curves of our program stars by using Maximum Entropy and Tikhonov regularization criteria will be carry out afterwards to derive area and temperature of active regions as well as their evolution over the years  (see, e.g. Lanza et al. \cite{Lanza06}). 
 
\subsection{Model}
We use the approach proposed by Dorren (\cite{Dorren87}) to model the amplitudes of the observed V magnitude, B$-$V and U$-$B color variations arising from the difference of fluxes in the U, B and V bands between opposite hemispheres of the active component. 
We take into consideration also the flux coming from the earlier-type component and, whenever it is the case, 
from a tertiary component. 
The stellar fluxes were determined by using the NextGen  atmosphere models of Hauschildt et al. (\cite{Hauschildt99}) for solar metallicity and convolved with the passbands of the UBV system (Johnson \cite{Johnson53}) as tabulated in Buser (\cite{Buser78}) and Buser \& Kurucz
 (\cite{Buserkurucz78}).
The adopted values of the components stellar effective temperature, radius and gravity
are listed in Sect. 2. 
The mentioned stellar fluxes and physical parameters were used to compute the total flux ratios between the cool and hot components in the U, B and V bands which are listed in Table\,\ref{target}, as well as to compute the magnitude variations arising from proximity and reflection effects according to Eq.\,(1) and (6) of Morris \& Naftilan (\cite{Morris93}). We assume that only the component whose total flux dominates the system's luminosity is active.
For instance, in the case of \object{BY Dra}, although both components have the same spectral class and should have similar levels of magnetic activity, we assume that only one component is active. In the case of \object{RS CVn}, the K0\,IV component, although less luminous than the F5\,IV component, will be considered the active one.
Limb-darkening coefficients, different for the unperturbed and the spotted photosphere, are taken from Diaz-Cordoves et al. (\cite{Diaz-Cordoves95}). 
We have computed the model magnitude and color variation for a grid of values of  temperature and covering fraction of spots and faculae.
The covering fraction was varied from 0. to 0.50 with a 0.01 increment, whereas the temperature of the surface inhomogeneities was varied from 3200 K up to 6800 K  with a 100\,K increment. For instance, we have computed our modelling also for a range of gravity values ($\Delta \log g \pm 1.5$) and effective temperature ($\Delta$T$_{\rm eff} \pm 150$) for sub-giant non-eclipsing binary stars in our sample, being their values poorly determined. 
In our modelling approach gravity-darkening effects are neglected by considering that these effects tend to cancel out when computing the flux difference between opposite hemispheres.  As reported in Sect.\,2,  the reflection effect is negligible for all program stars, whereas the proximity effect is marginal for \object{V711 Tau} and \object{AR Lac}. However, we notice that such an effect does not play any significant role in the observed color variation as verified by using Eq.\,(1) and (6) of Morris \& Naftilan (\cite{Morris93}).



\subsection{Results} 
\begin{figure}
\begin{minipage}{10cm}
\psfig{file=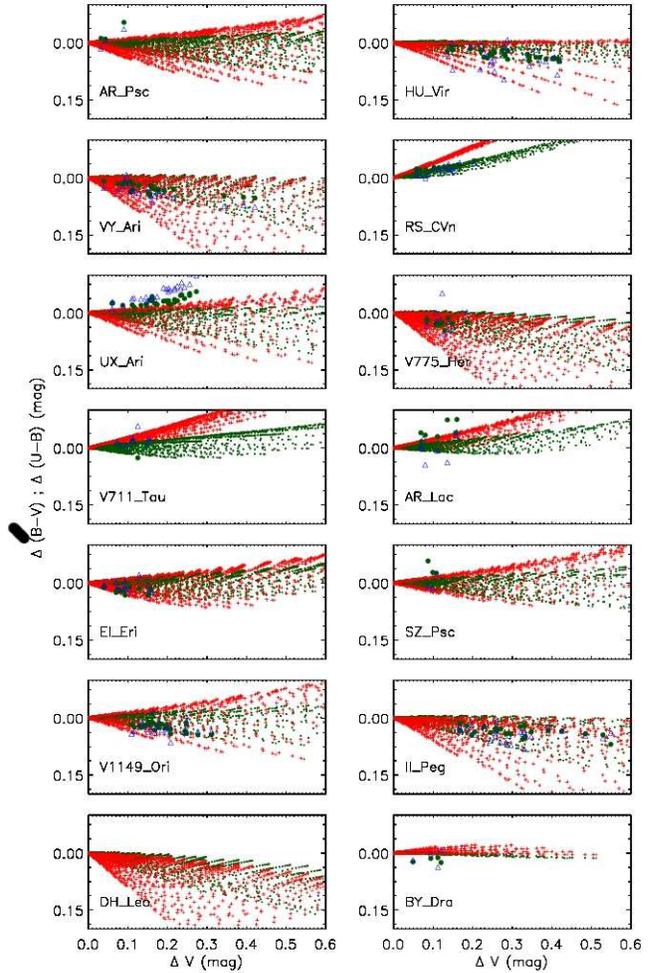,width=9cm}
\end{minipage}
\vspace{-1cm}
\caption{\label{fig-hot} Results of our model. Green filled bullets (B$-$V) and blue open triangles (U$-$B) present the observed magnitude and color variations. Green small dots (B$-$V) and  red small crosses (U$-$B) represent the family of model solutions.}
\end{figure}

In Fig.\,\ref{fig-hot} for each program star we plot the observed B$-$V (green filled bullets) and U$-$B (blue open triangles) vs. V magnitude variations and the model amplitudes (green small dots for B$-$V and red small crosses for U$-$B)  corresponding to all possible combinations of spot's temperature and area. Specifically, for a given spot temperature the model solutions for different values of filling factor dispose along a dotted-like curve. Dotted-like curves of different slopes correspond to different spot temperatures. 
As expected, model solutions are not unique: different combinations of temperature and filling factor can determine the same variation amplitudes. In general we found that the warmer the spots, the larger are their filling factor to fit the observed amplitude variations. Since our model is not able to obtain unique solutions, we will not focus on the found ranges of temperature and filling factor values, rather we shall discuss the possibility to fit the observed blueing by considering the contribution by earlier-type components. Here we present the results of the modelling for the individual program stars.


\noindent
\bf \object{AR Psc}: \rm  color variations are correlated ($\alpha < 0.1$) to V mag variations in 11\% of light curves. The model reveals that the observed blueing can be accounted in two out of three mean epochs, and within the photometric accuracy, by the presence of the earlier-type G5/6\,V component. In other words, the observed magnitude and color variations in two out of three cases fall within the area of models solutions.\\
\bf \object{VY Ari}: \rm  color variations are correlated ($\alpha < 0.1$) to V mag variations in 63\% of light curves. Consistently with its classification as spot-dominated star, the observed reddening can be attributed to the presence  of spots.\\
\bf \object{UX Ari}: \rm  color variations are correlated ($\alpha < 0.1$) to V mag variations in 83\% of light curves. The model reveals that the earlier-type G5\,V component cannot account for the observed blueing.\\
\bf \object{V711 Tau}: \rm color variations are correlated ($\alpha < 0.1$) to V mag variations in 29\% of light curves.  The model shows that the earlier-type G5\,V component can account for the observed blueing (except for one mean epoch).\\
\bf \object{EI Eri}: \rm  color variations are correlated ($\alpha < 0.1$) to V mag variations in 39\% of light curves. Consistently with its classification as spot-dominated star, the observed reddening can be attributed to the presence  of spots.\\
\bf V1149 Ori: \rm color variations are correlated ($\alpha < 0.1$) to V mag variations in 77\% of light curves. Notwithstanding the presence of the F8\,V component, this star does not show any evidence of blueing.  Consistently with its classification as spot-dominated stars, its color variations can be attributed to the only presence of spots. \\
\bf \object{DH Leo}: \rm since no color variations are found to be correlated ($\alpha < 0.10$) to V mag variations, we could not make a comparison with model solutions.\\
\bf \object{HU Vir}: \rm color variations are correlated ($\alpha < 0.1$) to V mag variations in  73\% of light curves. Consistently with its classification as spot-dominated star, the observed reddening can be attributed to the presence  of spots.\\
\bf \object{RS CVn}: \rm  color variations are correlated ($\alpha < 0.1$) to V mag variations in 95\% of light curves. The model shows that the flux contribution by the earlier-type F5\,V component can  account only for the B$-$V blueing, whereas the U$-$B color variations are systematically smaller than the model variations. \\
\bf \object{V775 Her}: \rm color variations are correlated ($\alpha < 0.1$)  to V mag variations in 38\% of light curves. Consistently with its classification as spot-dominated star, the observed reddening can be attributed to the presence  of spots.\\
\bf \object{AR Lac}: \rm  color variations are correlated ($\alpha < 0.1$) to V mag variations in 39\% of light curves. The model, with  flux contribution by the earlier-type G2\,IV component and spots, fits neither  the B$-$V blueing nor the U$-$B reddening. If fact, differently than expected in the case of spots, the B$-$V variations are systematically larger than U$-$B variations.\\
\bf \object{SZ Psc}: \rm color variations are correlated ($\alpha < 0.1$) to V mag variations in  37\% of light curves.  The model reveals that the earlier-type component cannot account for the observed blueing.\\
\bf \object{II Peg}: \rm color variations are correlated ($\alpha < 0.1$) to V mag variations in  69\% of light curves. Consistently with its classification as spot-dominated star, the observed reddening can be attributed to the presence  of spots.\\
\bf \object{BY Dra}:  \rm color variations are correlated ($\alpha < 0.1$) to V mag variations in 22\% of light curves.  Our model cannot fit the observed color variations, likely depending on the assuming that only one component is active and contributing to the observed color variations. \\
We note that among \it blueing \rm stars,  the presence of an earlier-type component can explain the observed blueing only in the case of \object{V711 Tau}.  As expected for all the \it reddening \rm stars, the spot model can fit the observed color variations in all mean epochs.\\

\section{Discussion}
The major result inferred from the analysis presented in Sect.\,4 is the existence of \it reddening \rm and \it blueing \rm stars which can be either \it color-correlated \rm or \it color-uncorrelated\rm.
As already anticipated, we guess that the existence of different patterns of color variation and different degrees of color-magnitude correlation can mainly arise from two circumstances: \it 1\,\rm) the presence of a fainter component of the binary system whose activity level is not negligible; \it 2\,\rm) the presence of hot faculae either spatially and temporally not correlated to cool spots. The latter represent, in turn, a strong observational evidence in favour of the existence of faculae at least in a few of our program stars. \\
\indent
Let us discuss the first circumstance. The activity level primarily depends on rotation rate and depth of the convection zone (see, e.g., Messina, Rodon\`o  \& Guinan \cite{Messina01}; Messina et al. \cite{Messina03}). Specifically, stars with shorter rotation period and deeper convection zone show photometric variability larger than slower-rotating and earlier spectral-type stars. For example, the activity level of the G5\,V component of \object{UX Ari} is much smaller than that of the K0\,IV component, because of the smaller depth of convection zone, although both components have the same rotation period. The luminosity difference between these components makes even more negligible the contribution by the G5\,V component to the observed system's variability. In order to quantify this contribution  by the earlier-type component, we have taken from the work of Messina et al. (\cite{Messina01}) the maximum light curve amplitude expected in the V band for the fainter active component of each program star. Using the $<b_{bv}>$ and $<b_{ub}>$ values from Table 6, although these values refer to the whole system, we have computed the approximate amplitude expected also in the B and U bands. Finally, considering the luminosity ratio between the cool and hot components as listed in Table 1, we have computed the maximum amplitude of the magnitude and color variations which could be attributed to the less active and fainter component.
We found that the SB1 stars \object{II Peg}, \object{VY Ari}, \object{HU Vir}, as well the SB2 stars V1149 Ori, \object{RS CVn}, whose earlier-type companion is inactive (spectral type earlier than F5\,V), or \object{UX Ari}, whose earlier-type companion if found with a negligible activity level, are all \it color-correlated \rm stars. In these systems the variability arises from only one component and the color-magnitude variation remains correlated. On the other hand, we found that for \object{V711 Tau}, \object{EI Eri}, \object{AR Lac}, \object{DH Leo} and \object{BY Dra} the variations arising from the fainter companion are significant, of the order of a few percents of magnitude. All these stars are \it color-uncorrelated \rm (the color-magnitude correlation is lowest for \object{BY Dra}, whose components have similar activity levels). We guess that since the surface inhomogeneities are distributed differently on both components, the respective patterns of color variations are not coherent and the correlation is more frequently lost. 
\object{AR Psc}, \object{V775 Her} and \object{SZ Psc} are the exceptions being \it color-uncorrelated\rm, although the variability contribution by the fainter companion is found to be negligible. The existence of faculae may play a major role in preventing the correlation in these stars. Indeed, we remind that \object{AR Psc} and \object{SZ Psc} were found to have $b_{ubv}$ negative and with very low significance level, respectively. \\
\indent
Let us discuss the second circumstance.
If we consider \it color-correlated \rm stars, we find that the slopes of the colors vs. the V mag variations are not constant vs. time. Since the slope value mostly depends on the average temperature of inhomogeneities, we deduce that it varies vs. time. Such variation may arise from the contemporary presence of inhomogeneities of different temperatures, e.g., cool spots and hot faculae, whose relative area and flux contributions are variable. \\
\indent
The existence of hot faculae in active stars is documented in a number of works.
Light curve inversion methods, widely used to extract information 
on the properties of stellar active regions, have generally assumed that dark spots are 
the dominant magnetic feature mainly responsible for the observed 
brightness variations. Indeed, it has been generally found that for the most active stars, i.e. stars with an 
activity level much higher than that of the Sun, neither faculae 
nor network elements are required to obtain quite satisfactory V-band light curve models, the effect of starspots
being dominant (e.g. Henry et al. \cite{Henry95}; Lanza et al. \cite{Lanza98}).
Since the 1990s, some authors began to realize that, for less active stars, the variability at optical wavelengths 
is significantly influenced, if not dominated,  by bright faculae. Radick et al. (\cite{Radick90}; \cite{Radick98}) found that the optical brightness of stars of solar age, or older, 
increases with increasing chromospheric activity level over time scales of several years, suggesting that such brightness enhancement may be mostly attributed to bright features. Radick et al. proposed a scenario according to which young and rapidly rotating stars arrange their surface magnetic flux predominantly into dark spots, whereas, when stars age and their rotation slows down, bright facula-like structures are favoured. However, the  rotational modulation of the optical flux remains dominated by dark spots in both young and old stars.\\
\indent
In a pilot program, Mirtorabi et al. (\cite{Mirtorabi03}) investigated the correlation between the optical light curve and the TiO absorption strength for the evolved chromospherically active star $\lambda$ And, and found clear evidence that the V-band and near-IR continua light variation primarily arise from bright rather than dark starspots. O'Neal et al. (\cite{ONeal98}) found, from spectroscopic data,  some evidence for multiple temperatures of the brightness inhomogeneities on \object{II Peg}. 
In the most active stars  faculae seem to be necessary to account for their UV excess with respect to inactive stars (e.g. Amado \cite{Amado03}).
The blueing of three stars in our sample, \object{UX Ari}, \object{V711 Tau} and \object{RS CVn}, has been previously investigated by Aarum Ulvas \& Engvold (\cite{Aarum03}) and Aarum Ulvas \& Henry (\cite{Aarum05}), and attributed to the presence of faculae.\\
\indent
We guess that there are epochs when spots and faculae are spatially associated
so that they can produce correlated magnitude and color variations.
Although area and/or temperature ratio can change, the correlation is anyway preserved.
There are other epochs in which spots and faculae are mostly spatially and temporally uncorrelated, e.g. spots and faculae have lifetimes significantly different and, when new spots or faculae emerge in the photosphere, their phase coherence with older patterns is lost. Although the global activity pattern
producing the V-band modulation is stable, that producing the color variation is less stable over the same timescale. In other words, during these epochs faculae seem to act as an interference source which destroys the correlation between color and magnitude variations arising from  spots only.\\
\indent
In our sample the \it reddening \rm stars \object{EI Eri}, \object{V775 Her} and \object{DH Leo} and the \it blueing \rm stars \object{SZ Psc} and \object{AR Psc}, whose fainter components were shown to give no contribution to the observed variability, are the best candidates for hosting faculae which are most of the time uncorrelated with spots. Although the correlation between spots and faculae is absent on the rotational time scale, it is still present on the longer time scale. In fact, as shown in the bottom panels of Fig.\,2 and 7, as far as these stars approach the maximum activity level and the total amount of spots increases (making the star fainter), also the total amount of faculae increases (making the star bluer). However, either  slope and correlation coefficient are much smaller than in \it color-correlated \rm stars. \\
Also the remaining \it blueing \rm stars \object{AR Lac}, \object{RS CVn}, and \object{UX Ari} are the best candidates for hosting faculae, since their earlier-type
stellar component cannot account for the observed blueing, according to the results of our modelling.

\section{Conclusions}
The long-term monitoring project of active close binary systems carried out at OAC has allowed us to collect a time-extended database of multiband high-precision photometric observations for a sample of 14 program stars. Correlation, regression analyses, as well as simple modelling approach for these data has allowed us to discover and interpret the existence of color vs. magnitude variations showing different patterns and level of correlations. Here we report the most relevant conclusions of our study: \\

\begin{itemize}
\item[] \it General conclusions:\rm
\item 
Active close binary systems show magnitude and color variations. Such variations are
correlated to each other in a few seasons, whereas the correlation is lost in other seasons.
The correlation is found more frequently (in more that 60\% of the observed seasons) in single- and double-lined stars in which only one component is active, the other component being inactive (F spectral type or earlier) or with a negligible activity level. The correlation is much less frequent (in less that 40\% of observed seasons) in  double-lined stars whose fainter component has a non-negligible level of activity. We may guess that in these stars the correlation is prevented by different spot distributions on the two active components. 
\item
The slope of the relation between magnitude and color variations is found to vary from season to season in all stars. Such variation may arise from a variable flux contribution by contemporary present inhomogeneities of different temperature, such as cool spots and hot faculae, which make variable the average temperature and, therefore, the computed relation slope.
\item
In single- and double-lined stars in which only one component is active the correlation between magnitude and color variation is found in a few seasons to be absent likely because spots and faculae are spatially and temporally uncorrelated. In such cases faculae play like a noise source which 
makes  the observed color pattern highly unstable. In double-lined stars where the fainter component has a significant level of activity, 
beside faculae, also a different inhomogeneity pattern on the fainter component, can prevent the correlation between magnitude and color variation.
\item 
Our modelling shows that for \it blueing \rm stars an earlier-type component can give a significant contribution to the observed blueing. However, it cannot account alone for the variable slope of the magnitude-color relations.\\

\item[]\it Specific conclusions:\rm
\item
\object{II Peg}, \object{VY Ari}, \object{HU Vir} and V1149 Ori are \it reddening \rm and \it color-correlated \rm  stars whose variability originates only from the primary component and is dominated by cool spots. However,  faculae can be present and vary from time to time the reddening slope, but preserving in most seasons their correlation with spots;\\
\object{V775 Her} is  a \it reddening \rm and \it color-uncorrelated \rm star whose variability originates only from the primary component and it arises from either spots or facular activity not correlated to spots;\\
\object{BY Dra}, \object{EI Eri} and \object{DH Leo} are \it reddening \rm and \it color-uncorrelated \rm  stars whose
variability originates from both components and is dominated by cool spots.  The correlation between their magnitude and color variations is frequently lost due to either non-negligible activity level in the fainter companion or presence of faculae non-correlated to spots;\\
{\bf \object{UX Ari} and \object{RS CVn} are  {\it blueing}  and {\it color-correlated}  stars whose activity takes place  only in the primary component. Their short-term color variations are dominated by faculae, which are most of the time correlated to spots. The long-term color variation can partly arise form the hotter companion};\\
\object{AR Psc} and \object{SZ Psc} are \it blueing \rm  and \it color-uncorrelated \rm  stars  whose activity takes place only in the primary component and whose color variations are dominated by faculae which are on the rotation timescale rarely correlated to spot activity;\\
\object{AR Lac} and \object{V711 Tau} are \it blueing \rm and \it color-uncorrelated \rm  stars  whose activity takes place in both components and whose color variations are dominated by faculae. These are rarely correlated to magnitude variation due to either non-correlated spot/faculae activity on the rotation timescale and a non-negligible activity level in the fainter companion. \object{V711 Tau} is the only star in our sample whose blueing, but not the slope variations, could be entirely attributed to the flux contribution by the earlier-type component.

\end{itemize}

\begin{acknowledgements}
The acquisition of photometric data over so many years with the Catania APT has been possible thanks to the dedicated 
and highly competent technical assistance of a number of people, notably P.~Bruno, E.~Martinetti and S.~Sardone. 
Active star research at the INAF-Catania Astrophysical Observatory is funded by MUR ({\it Ministero dell'Universit\`a e della Ricerca}), and the {\it 
Regione Sicilia}, whose financial support is gratefully acknowledged.
The extensive use of the SIMBAD and ADS databases, 
operated by the CDS center,
(Strasbourg, France), is also gratefully acknowledged. I would like to thank Dr. A.F. Lanza for his valuable comments and useful discussion, and the Referee, Dr. Panos G. Niarchos, for careful reading of the manuscript.
\end{acknowledgements}

\Online


\begin{deluxetable}{c l l l l l l l l l l l}
\tablecolumns{12}
\tablecaption{\label{tab-photom} { Summary of photometric observations: mean epoch of observation, mean, initial and final heliocentric Julian day, number of points in the light curve, brightest V magnitude and light curve peak-to-peak amplitude in the V band, B$-$V and U$-$B colors, standard deviation ($\sigma_v$) of v$-$c and ($\sigma_{ck1-c}$) of ck1$-$c differential observations. The telescopes are labelled with 'A': Phoenix-25, 'B': APT80/1, 'C': ESO 50cm.}}
\small
\tablewidth{0pt}
 \tablehead{
\colhead {Mean Epoch}   & \colhead {HJD$_{\rm mean}$} & \colhead {HJD$_{\rm ini}$} & \colhead {HJD$_{\rm end}$ } & \colhead {N$_{\rm m}$}   & \colhead {V$_{\rm min}$} & \colhead {$\Delta$V} & \colhead { B$-$V} &
 \colhead {U$-$B}      & \colhead { $\sigma_v$ }       & \colhead {$\sigma_{ck1-c}$}  & \colhead {  Note} \\    
 }
\startdata
\multicolumn{12}{c}{\bf \object{AR Psc}} \\
 1987.93 & 7136.63 & 7129.64 & 7143.61 & 13 & 7.243 & 0.072 & 0.855 & 0.392 & 0.026 & 0.007 & C\\
 1989.96 & 7879.09 & 7870.61 & 7887.57 & 13 & 7.273 & 0.056 & 0.856 & 0.380 & 0.022 & ---   & C \\
 1990.83 & 8195.35 & 8178.90 & 8211.80 & 14 & 7.299 & 0.091 & 0.845 & 0.393 & 0.033 & 0.007 & A  \\
 1991.82 & 8556.83 & 8540.87 & 8572.78 & 11 & 7.343 & 0.065 & 0.837 &  ---- & 0.021 & 0.006 & A  \\
 1992.00 & 8621.69 & 8589.79 & 8653.60 & 11 & 7.302 & 0.107 & 0.843 & 0.387 & 0.039 & 0.006 & A  \\
 1992.85 & 8934.10 & 8899.83 & 8968.37 & 20 & 7.312 & 0.056 & 0.840 & 0.398 & 0.018 & 0.005 & A+B  \\
 1993.02 & 8994.98 & 8978.36 & 9011.61 & 17 & 7.290 & 0.094 & 0.856 & 0.388 & 0.031 & 0.005 & B  \\
\enddata
\end{deluxetable}
\clearpage

 \begin{deluxetable}{r@{\hspace{.01cm}} r@{\hspace{.15cm}} r@{\hspace{.15cm}} r@{\hspace{.15cm}} c@{\hspace{.01cm}} r@{\hspace{.15cm}} r@{\hspace{.15cm}} r@{\hspace{.15cm}} c@{\hspace{.01cm}} r@{\hspace{.15cm}} r@{\hspace{.15cm}} r@{\hspace{.15cm}} c@{\hspace{.01cm}}}
\tablecolumns{13}
\tablecaption{{ Summary of regression and correlation analyses: mean epoch of light curve, number of data points, slopes $b_{bv}$, $b_{ub}$ and $b_{ubv}$ along with correlation coefficients and significance levels.\label{coeff-rot}} }
\small
\tablewidth{0pt}
 \tablehead{
\colhead {HJD$_{\rm mean}$}   & \colhead {N$_V$} & \colhead {b$_{bv}$} & \colhead {r$_{bv}$ } & \colhead {$\alpha$}   & \colhead {N$_{BV}$} & \colhead {b$_{ub}$} & \colhead {r$_{ub}$ } &
 \colhead {$\alpha$}      & \colhead {N$_{UBV}$} & \colhead {b$_{ubv}$} & \colhead {r$_{ubv} $} &
 \colhead {$\alpha$}   \\
 }
\startdata
\multicolumn{13}{c}{\bf \object{AR Psc}} \\
   7136.6258 &  13 &  0.025 &  0.188 &         $\alpha>$0.2 &  13 &  0.155 &  0.188 &         $\alpha>$0.2 &  13 &  0.225 &  0.103 &         $\alpha>$0.2 \\
   7879.0872 &  13 & $-$0.077 & $-$0.351 &         $\alpha>$0.2 &  13 & $-$0.128 & $-$0.351 &         $\alpha>$0.2 &  13 & $-$0.641 & $-$0.547 &     0.1$<\alpha<$0.2 \\
   8195.3506 &  14 & $-$0.066 & $-$0.259 &         $\alpha>$0.2 &  14 & $-$0.165 & $-$0.269 &         $\alpha>$0.2 &  13 & $-$0.593 & $-$0.324 &         $\alpha>$0.2 \\
   8556.8265 &  11 & $-$0.332 & $-$0.722 &        $\alpha<$0.05 & --- & --- & --- & --- & --- & --- & --- & --- \\
   8621.6923 &  11 &  0.021 &  0.179 &         $\alpha>$0.2 &  11 & $-$0.233 &  0.304 &         $\alpha>$0.2 &   6 &  0.684 &  0.290 &         $\alpha>$0.2 \\
   8934.0984 &  20 & $-$0.016 & $-$0.059 &         $\alpha>$0.2 &  20 &  0.028 & $-$0.059 &         $\alpha>$0.2 &  20 & $-$0.327 & $-$0.132 &         $\alpha>$0.2 \\
   8994.9811 &  17 &  0.024 &  0.151 &         $\alpha>$0.2 &  17 &  0.083 &  0.151 &         $\alpha>$0.2 &  17 &  0.676 &  0.392 &         $\alpha>$0.2 \\
\enddata
\end{deluxetable}

\begin{table*}
\caption{{ Average slopes and related uncertainty of the linear fits to:  B$-$V vs. V, U$-$B vs. V and U$-$B vs. B$-$V relations. Only light curves (N) whose magnitude and colors were correlated with a high significance level ($\alpha < 0.1$) are considered. The slopes' smallest and largest values are also listed.}}\label{tabcoeff_short}
\begin{center}
\begin{tabular}{llrrrr}
\hline
\multicolumn{6}{c}{}\\
\multicolumn{6}{c}{\bf Short$-$term rotational modulation}\\
\multicolumn{6}{c}{}\\
\hline
Target        &  & N        &  Average  Slope     & Min     & Max        \\
\hline    
\object{AR Psc}           &  $<$BV/V$>$ &   4 &  -0.39$\pm$  0.15 &  -0.60 &  -0.23 \\
         &  $<$UB/V$>$ &   4 &  -0.12$\pm$  0.41 &  -0.40 &   0.47 \\
         &  $<$UB/BV$>$ &   7 &  -0.70$\pm$  0.31 &  -1.16 &  -0.15 \\
	 \hline
\object{VY Ari}           &  $<$BV/V$>$ &  24 &   0.18$\pm$  0.07 &   0.09 &   0.35 \\
\end{tabular}
\end{center}
\end{table*}

\begin{longtable}{lcrrr}
\caption{{ Slopes and related uncertainty of the linear fits to: B$-$V vs. V, U$-$B vs. V and U$-$B vs. B$-$V relations on the long-term time scale, to the light curve brightest magnitude vs.  bluest colors ($_{\rm min}$) and faintest magnitude vs. reddest colors ($_{\rm max}$). Number N of observations, correlation coefficient (r) and related significance level are also listed.}}\label{tabcoeff_long}\\
\hline
\multicolumn{5}{c}{}\\
\multicolumn{5}{c}{\bf Long$-$term variation}\\
\multicolumn{5}{c}{}\\
\hline
           &      N     & Slope     &  r     &  Significance \\
\hline
\endfirsthead
\caption{Continued. }\\
\hline
           &      N     & Slope     &  r     &  Significance \\
\hline\hline
\endhead
\endlastfoot
\multicolumn{5}{c}{\bf AR Psc }\\
                B$-$V/V & 595 &  $-$0.08$\pm$  0.01 &  $-$0.27 &      $\alpha < 0.01$ \\
            (B$-$V/V)$_{\rm min}$ &  34 &  $-$0.08$\pm$  0.05 &  $-$0.28 & $0.05< \alpha < 0.1$ \\
            (B$-$V/V)$_{\rm max}$ &  34 &  $-$0.10$\pm$  0.05 &  $-$0.34 &      $\alpha < 0.01$ \\
                          U$-$B/V &  558 &  $-$0.11$\pm$  0.01 &  $-$0.37 &      $\alpha < 0.01$ \\
            (U$-$B/V)$_{\rm min}$ &  33 &  $-$0.10$\pm$  0.04 &  $-$0.38 &      $\alpha < 0.01$ \\
            (U$-$B/V)$_{\rm max}$ &  33 &  $-$0.13$\pm$  0.04 &  $-$0.56 &      $\alpha < 0.01$ \\
                         U$-$B/B$-$V & 558 &  $-$0.00$\pm$  0.04 &  $-$0.00 &      $\alpha < 0.01$ \\
\hline
\end{longtable}

\end{document}